\documentclass[prd,aps,11pt,nofootinbib,superscriptaddress,preprintnumbers,balancelastpage,longbibliography]{revtex4-1}

\usepackage{amsmath}
\usepackage{amsfonts}
\usepackage{amssymb}
\usepackage{textcomp}
\usepackage{graphicx, rotating}
\usepackage{epstopdf}
\usepackage{epsfig}
\usepackage{latexsym}
\usepackage{graphicx}
\usepackage{color}
\usepackage{amsmath,amssymb}
\usepackage{slashed}
\usepackage[vcentermath]{youngtab}

\usepackage[colorlinks]{hyperref}
\hypersetup{
    colorlinks=true,
    linkcolor=rossos,
    filecolor=blue,      
    urlcolor=bluscuro,
    citecolor=bluscuro,
}

\usepackage{multirow}
\usepackage{stackengine}

\usepackage[pagewise]{lineno} 

\usepackage[utf8]{inputenc}

\graphicspath{ {./Figures/} }

\definecolor{rossos}{cmyk}{0,1,1,0.55}
\definecolor{bluscuro}{rgb}{0.15, 0.2, .85}
\definecolor{bluchiaro}{cmyk}{1,.3,0.,0.1}
\definecolor{green1}{rgb}{0.0, 0.5, .0}

\newcommand{\be}{\begin{equation}}
\newcommand{\ee}{\end{equation}}
\newcommand{\bea}{\begin{eqnarray}}
\newcommand{\eea}{\end{eqnarray}}
\newcommand{\bc}{\begin{center}}
\newcommand{\ec}{\end{center}}

\def\tr{{\rm tr\,}}

\definecolor{red}{rgb}{1,0,0}
\definecolor{pink}{rgb}{1,0,1}

\numberwithin{equation}{section}

\begin{document}

\title{On the Proof of Chiral Symmetry Breaking from Anomaly Matching in QCD-like Theories}

\author{Luca Ciambriello}
\email{luca.ciambriello@unicatt.it}
\affiliation{\small Interdisciplinary Laboratories for Advanced Materials Physics (i-LAMP) and Dipartimento di Matematica e Fisica, Universit\`{a} Cattolica del Sacro Cuore, Brescia, Italy}

\author{Roberto Contino}
\email{roberto.contino@uniroma1.it}
\affiliation{\small Dipartimento di Fisica, Sapienza Universit\`{a} di Roma, Italy}
\affiliation{\small Istituto Nazionale di Fisica Nucleare (INFN), Sezione di Roma, Italy}

\author{Andrea Luzio}
\email{andrea.luzio@sns.it}
\affiliation{\small Scuola Normale Superiore, Pisa, Italy}
\affiliation{\small Istituto Nazionale di Fisica Nucleare (INFN), Sezione di Pisa, Italy}

\author{\\Marcello Romano}
\email{marcello.romano@ipht.fr}
\affiliation{\small Université Paris-Saclay, CNRS, CEA, Institut de Physique Théorique, 91191, Gif-sur-Yvette, France}

\author{Ling-Xiao Xu}
\email{lxu@ictp.it}
\affiliation{\small Dipartimento di Fisica e Astronomia `G. Galilei', Universit\`{a} di Padova, Italy}
\affiliation{\small Istituto Nazionale di Fisica Nucleare (INFN), Sezione di Padova, Italy}
\affiliation{\small Abdus Salam International Centre for Theoretical Physics, Trieste, Italy}


\begin{abstract}
\vspace{0.5cm}
We critically reconsider the argument based on 't Hooft anomaly matching that aims at proving chiral symmetry breaking in confining four-dimensional QCD-like theories with $N_c>2$ colors and $N_f$ flavors.
The main line of reasoning relies on a property of the solutions of the anomaly matching and persistent mass equations called $N_f$-independence. In previous works, the validity of $N_f$-independence was assumed based on qualitative arguments, but it was never proven rigorously. We provide a detailed proof and clarify under which (dynamical) conditions it holds.
Our results are valid for a generic spectrum of massless composite fermions including baryons and exotics.
\end{abstract}

\maketitle
\newpage 
{\small\tableofcontents}

\section{Introduction}\label{sec:intro}

Confinement and chiral symmetry breaking are the two phenomena exhibited by QCD at low energy. Their full theoretical understanding still remains elusive due to the fact that QCD gets strongly coupled in the infrared (IR). One particular question is whether confinement enforces chiral symmetry breaking inevitably in QCD.
The common lore is that a phase of confinement without chiral symmetry breaking does not occur in QCD-like theories with $SU(N_c)$ gauge group and $N_f$ flavors of vectorlike quarks in the fundamental representation. (We clarify our working definition of confinement and how we use it in Section~\ref{sec:spectrum}.) This must be contrasted with ${\cal N}=1$ supersymmetric QCD, where such a phase, dubbed s-confinement, is known to exist for $N_f = N_c+1$~\cite{Seiberg:1994bz}.
In this work, we address this question by critically reconsidering a class of arguments based on 't Hooft anomaly matching~\cite{tHooft:1979rat}.
A QCD-like theory is expected to be in the confining regime when the number of massless flavors is smaller than a critical value $N_f^{CFT}$, above which it is either conformal (for $N_f^{CFT}\leq N_f < 11N_c/2 $) or IR free (for $N_f \geq 11 N_c/2$). Therefore, we consider $N_c>2$ and $2\leq N_f<N_f^{CFT}$. For $N_c=2$, quarks and antiquarks transform in equivalent color representations, and this case needs to be addressed separately.

Among the available theoretical tools, 't Hooft anomaly matching stands out as one of the few that can constrain the IR spectrum of confining gauge theories. A QCD-like theory with $N_f$ massless flavors has a global chiral symmetry~\footnote{In this paper, we neglect the discrete factors that appear in the global symmetry group. We define $U(1)_B$ in a standard way and assign baryon number $1/N_c$ to quarks.} 
\bea
\mathcal{G}[N_f]=SU(N_f)_L\times SU(N_f)_R \times U(1)_B\, ,
\label{def:G}
\eea 
where $U(1)_B$ is baryon number.
If chiral symmetry is unbroken by the vacuum, the color-singlet bound states in the low-energy spectrum
transform in non-trivial irreducible representations (irreps) of $\mathcal{G}[N_f]$. Furthermore, massless composite fermions must exist to reproduce in the infrared (IR) the anomalies of $\mathcal{G}[N_f]$ induced by the quarks in the ultraviolet (UV). In other words, the UV and IR anomalies of the chiral symmetry group $\mathcal{G}[N_f]$ need to be matched. This implies a set of equations called Anomaly Matching Conditions (AMC). They were first derived by 't Hooft in Ref.~\cite{tHooft:1979rat} by coupling background gauge fields to the anomalous quark currents and introducing color-neutral spectator fermions. A derivation based on analyticity and unitarity was given in Refs.~\cite{Frishman:1980dq,Coleman:1982yg}, see also Ref.~\cite{Coleman:1980mx} for implications in the large-$N_c$ limit.

Besides 't Hooft anomaly matching, the low-energy spectrum of QCD-like theories also needs to satisfy Persistent Mass Conditions (PMC)~\cite{Preskill:1981sr}, originally formulated by 't Hooft as decoupling conditions~\cite{tHooft:1979rat}. As we will see, PMC are an implication of the Vafa-Witten theorem valid for vectorlike gauge theories with $\theta=0$~\cite{Vafa:1983tf}.
Their physical significance can be summarized as follows: when $i$ flavors are given finite and unequal masses, with $1\leq i\leq N_f-2$, the symmetry $\mathcal{G}[N_f]$ is explicitly broken to 
\bea
\mathcal{G}[N_f,i]=SU(N_f-i)_L\times SU(N_f-i)_R \times U(1)_{H_1} \times \cdots \times U(1)_{H_i} \times U(1)_B\, ,
\label{def:G_prime}
\eea
where each vector factor $U(1)_{H_k}$ is defined such that the $k$-th massive quark has charge $H_k=1$, while the other quarks have $H_k=0$.
Bound states that are charged under $U(1)_{H_1} \times \cdots \times U(1)_{H_i}$ must contain massive flavors of quarks or antiquarks as their microscopic constituents, and the unbroken chiral symmetry cannot prevent them from acquiring a mass. In other words, such composite fermions with non-zero charge under $U(1)_{H_1} \times \cdots \times U(1)_{H_i}$ must furnish vectorlike representations under~$\mathcal{G}[N_f,i]$.~\footnote{Here we focus on bound states with non-zero baryon number, which therefore transform as complex representations of  $\mathcal{G}[N_f,i]$. Only Dirac mass terms are allowed for these states.} Mathematically, this translates into the PMC. 
In particular, PMC equations with more than one massive flavors were not fully exploited in the previous literature but, as it turns out, they are key in the approach of Ref.~\cite{Ciambriello:2024xzd}.

One possible strategy to prove chiral symmetry breaking consists in making use of AMC and PMC. These form a system of linear equations in the indices of massless composite fermions. The index of a given irrep of $\mathcal{G}[N_f]$ is defined as the number of left-handed minus right-handed helicity states in the spectrum transforming as such irrep. If there exist no integral values of these indices that satisfy both AMC and PMC, then chiral symmetry breaking must occur. This line of reasoning was first proposed by 't~Hooft~\cite{tHooft:1979rat} and afterwards pursued by many other authors~\cite{Frishman:1980dq, Schwimmer:1981yy, Farrar:1980sn, Takeshita:1981sx, Kaul:1981fd, Cohen:1981iz,Bars:1981nh} under specific assumptions.
In his seminal paper, 't Hooft worked out the cases $N_c=3$ and $N_c=5$, showing that no integral solution exists for $N_f>2$. He assumed that the spectrum of massless fermions includes only baryons with baryon number $b=1$, and that mixed representations have vanishing index for $N_c=5$.
Frishman et al.~\cite{Frishman:1980dq} extended the analysis to generic $N_c$ by working under the same assumptions. They found no integral solution. (The real solution found in Ref.~\cite{Frishman:1980dq} is one where `elbow'-shape Young Tableaux have indices $\pm1/N_c^2$, while all the other indices are zero. For $N_c=5$, this coincides with the real solution found by 't Hooft.)
Considering only massless baryons, Cohen and Frishman~\cite{Cohen:1981iz} further showed that integer solutions do exist for $N_f=2$ and arbitrary $N_c$ besides the $\sigma$-model solution already found by 't Hooft.

An important step forward towards a general proof was made by Farrar in Ref.~\cite{Farrar:1980sn}. She considered the possible existence of exotic massless bound states in the spectrum, i.e. bound states with antiquark constituents, and was able to prove chiral symmetry breaking by assuming `$N_f$-independence'. According to the latter, the solutions of AMC and PMC are the same for any number of flavors. This property was first proposed by 't Hooft in Ref.~\cite{tHooft:1979rat} and justified by him through qualitative dynamical considerations. It was later suggested by Ref.~\cite{Frishman:1980dq} that it comes as a consequence of the PMC, and similar arguments were presented in a more quantitative fashion in~\cite{Takeshita:1981sx,Kaul:1981fd}. On the other hand, Cohen and Frishman pointed out in Ref.~\cite{Cohen:1981iz} that, if only minimal baryons are considered as massless bound states, it is necessary to separate the regimes with $N_f>N_c$ and $N_f\leq N_c$. Their work suggests that $N_f$-independence cannot hold in general, but only for sufficiently large $N_f$ and according to the spectrum of putative massless composite fermions. 

A different argument to prove chiral symmetry breaking using 't~Hooft anomaly matching was pursued independently by Schwimmer in Ref.~\cite{Schwimmer:1981yy}. He considered an $SU(N_f|N_f)$ superalgebra whose grading acts so that odd supergenerators mix massless states with opposite helicity. Although strictly speaking $SU(N_f|N_f)$ is not a symmetry of a QCD-like theory, it can be used as a classification group. Schwimmer argued that, for a purely baryonic massless spectrum, irreducible representations of $SU(N_f|N_f)$ satisfy PMC trivially, but they cannot furnish an integral solution of AMC.
In order to elevate this simple and elegant argument to the level of a proof of chiral symmetry breaking, one should demonstrate that it works for a generic spectrum with exotics and that it gives the most general solution of PMC, i.e. that every PMC solution is a collection of superalgebra irreps. This evidence is still missing.

Despite many papers on the subject and the various approaches explored in the literature, proving chiral symmetry breaking in a generic QCD-like confining theory is still an open issue. In particular, one would like a proof valid for the most generic spectrum of massless bound states and for arbitrary $N_c$ and $N_f$.
In this work, we reconsider the argument based on 't Hooft anomaly matching and $N_f$-independence with the aim of formulating it in a rigorous fashion.
We derive a proof of $N_f$-independence as a mathematical property of AMC and PMC equations and clarify under which assumptions it holds. The latter turn out to be dynamical requirements on the spectrum of massless bound states. Our study extends previous analyses and corrects a few statements on the form of the solution that were made in the literature. We report in a companion paper~\cite{Ciambriello:2024msu} the explicit solutions of the system of AMC and PMC equations that we derived for specific values of $N_c$ and $N_f$, and focus here on the more formal arguments. In a third paper, we further provide two novel arguments to prove chiral symmetry breaking which apply to a generic spectrum of massless bound states~\cite{Ciambriello:2024xzd}.

There exist other arguments, not based on 't Hooft anomaly matching, that can give evidence to the occurrence of chiral symmetry breaking in the confining description of QCD-like theories.
One attempt makes use of the mass inequality $m_B\geq m_\pi$, which has been proven by Weingarten to hold in QCD-like theories~\cite{Weingarten:1983uj}, see also~\cite{Nussinov:1999sx}. Here $m_B$ is the mass of the lightest baryon and $m_\pi$ is the mass of the lightest state interpolated by the operator $\bar q \gamma^5 q$. If one assumes that chiral symmetry is not spontaneously broken in the confining description, then matching 't Hooft anomalies requires $m_B=0$, and the mass inequality implies $m_\pi=0$. Although this proves that there must exist at least one massless scalar in the spectrum (something which is in fact rather interesting per se), it does not contradict the initial assumption of unbroken chiral symmetry, since the axial current $J^\mu_A$ and the state~$\pi$ transform differently under $SU(N_f)_L\times SU(N_f)_R$, hence $\langle 0 | J^\mu_A| \pi \rangle = 0$. Therefore, this argument does not seem conclusive.
A different approach aims at deducing the properties of real QCD from those of $\mathcal{N}=1$ SUSY QCD~\cite{Seiberg:1994bz,Seiberg:1994pq} by adding small supersymmetry-breaking terms~\cite{Aharony:1995zh,Cheng:1998xg,Arkani-Hamed:1998dti,Luty:1999qc,Abel:2011wv,Murayama:2021xfj,Luzio:2022ccn}. Although the IR dynamics of near-SUSY QCD can be characterized rather precisely and various exotic phases can be identified, extrapolating to real QCD by sending the supersymmetry-breaking scale to infinity is challenging due to the possible occurrence of phase transitions~\cite{Luzio:2022ccn,Dine:2022req}. In this work, we do not rely on supersymmetry.

The paper is organized as follows.
\begin{itemize}
\item In Section~\ref{sec:spectrum} we clarify what we mean by confinement and characterize the most general spectrum of massless composite fermions using tensor notation. Massless fermions can be classified by their representations under the chiral symmetry group ${\cal G}[N_f]$, and are interpolated from the vacuum by local composite operators which transform as tensors of ${\cal G}[N_f]$.

\item Section~\ref{sec:overview} gives a concise review on AMC and PMC equations. In particular, we justify PMC from the Vafa-Witten theorem~\cite{Vafa:1983tf}, and make some original considerations regarding PMC with more than one massive flavor. 

\item We enunciate and prove our theorem on $N_f$-independence in Section~\ref{sec:Nf-independence}. There we derive several features and important details that were overlooked in the previous literature, in particular the conditions under which $N_f$-independence holds true. 

Let us define `class A' tensors as composite operators made of $n_q$ quarks and $n_{\bar q}$ antiquarks (where $n_q= b N_c+n_{\bar q}$, and $b$ is the baryon number) satisfying $N_f> b N_c + 2\, n_{\bar q}$.
We find that $N_f$-independence is valid when the only massless bound states to have non-vanishing indices are those interpolated by `class A' tensors. For the baryonic case, our result is consistent with Ref.~\cite{Cohen:1981iz} when $b=1$ and $n_{\bar q}=0$.

\item Finally, we conclude in Section~\ref{sec:conclusion} and clarify the most technical aspects of our analysis in Appendices~\ref{app:trace},~\ref{app:eq_tensor},~\ref{app:PMCsamemasses} and~\ref{app:exotics}. 
\end{itemize}
Throughout the paper, several examples are provided which we hope can be useful for the reader.

\section{Characterizing Massless Bound States}
\label{sec:spectrum}

A precise definition of confinement is notoriously difficult in theories with quarks in the fundamental representation of $SU(N_c)$ gauge group (see e.g.~\cite{Greensite:2011zz}). In pure Yang-Mills theories, confinement is associated with a linear potential between two test charges, and can be characterized by means of a non-local order parameter and the center symmetry.
In modern terminology, the latter is referred to as a one-form symmetry~\cite{Gaiotto:2014kfa}.
For QCD-like theories with massive quarks in the fundamental representation, the center symmetry can be viewed as an accidental symmetry emerging at energies below the lightest quark mass. 
Nevertheless, none of these tools applies to QCD-like theories with massless quarks in the fundamental representation, where the confining string is screened even at low energy, hence the center symmetry is fully broken. 

A related wisdom is that there is no sharp distinction between the confining regime and the Higgs regime for theories without center symmetry, a phenomenon usually dubbed as Higgs-confinement continuity (see e.g.~\cite{Dumitrescu:2023hbe} for a recent discussion using modern terminology). This suggests that there may be a smooth deformation from one regime of the theory to the other without encountering a phase transition, hence both the confining and Higgs regimes actually belong to the same phase according to the Landau paradigm. 

In this paper, by `confinement' we mean a low energy effective description of the QCD-like theory, where particles (hadrons) are interpolated by color-singlet local operators. We assume, in particular, that there exist no massless extended objects interpolated by non-local operators. The notion of particles implies that the low energy effective theory flows to an infrared free fixed point at zero energy. A similar confining description is realized in supersymmetric QCD~\cite{Seiberg:1994bz} and other supersymmetric gauge theories~\cite{Csaki:1996sm,Csaki:1996zb}, and is usually dubbed as s-confinement~\cite{Intriligator:1995au}.~\footnote{Other terms such as `color confinement'~\cite{Greensite:2011zz}, `quark confinement'~\cite{Shifman:2012zz, Schwartz:2014sze}, `fermion trapping'~\cite{Weinberg:1996kr} and `screening confinement'~\cite{Csaki:1996zb} are also commonly used in the literature.} Even though such a low energy effective theory per se does not define a phase in the Landau sense (due to the lack of center symmetry), it is still a valid description when the coupling between quarks and gluons becomes very strong. 
We will study whether such a confining description is compatible with unbroken chiral symmetry in QCD-like theories.

\subsection{Massless Bound States and Composite Operators}
We focus on values of $N_c$ and $N_f$ for which the theory is in the confining regime, and assume that chiral symmetry is unbroken. This requires the existence of massless fermionic bound states to reproduce 't Hooft anomalies in the IR.  According to the Weinberg-Witten theorem~\cite{Weinberg:1980kq}, a theory with Lorentz-covariant conserved currents cannot contain massless particles of spin $j> 1/2$ that have a non-vanishing corresponding charge. Therefore, we only consider fermionic bound states of spin $j=1/2$. 
Massless bound states in the spectrum can be classified in terms of irreducible representations (irreps) of the flavor symmetry group $\mathcal{G}[N_f]$, cf.~Eq.~(\ref{def:G}). In particular, one can study the gauge-invariant composite local operators that interpolate massless fermions from the vacuum. Being color singlets, such operators are characterized by the sum rule
\bea
\label{eq:cons}
n_{q} - n_{\bar q} = b N_c\ ,
\eea
where $n_q$, $n_{\bar q}$ are the numbers of quark and antiquark fields forming the operator, and $b$ is its baryon number.
Since the bound state is a fermion, $b$ and $N_c$ must be both odd integers. As explained in the next section, one can focus on bound states with positive~$b$ without loss of generality.

We define \emph{baryons} those bound states that can be interpolated by operators with only quarks (i.e. with $n_{\bar q} =0$), and call \emph{exotics} all the other bound states. Exotics thus transform in representations of ${\cal G}[N_f]$ that cannot be obtained through operators with only quarks.

As a consequence of their quark and antiquark content, composite operators transform as tensors of $SU(N_f)_L\times SU(N_f)_R$ with $n_{q}$  contravariant upper indices and $n_{\bar q}$ covariant lower indices. One can thus establish a correspondence between the space of interpolating operators and the abstract space of tensors of $SU(N_f)_L\times SU(N_f)_R$.
Notice that, while a given operator corresponds to a single tensor, the converse is not true: different operators with the same quark and antiquark (hence flavor) content will correspond to the same tensor.
Traceless tensors~\footnote{A tensor is defined to be traceless if the ($SU(N_f)_L\times SU(N_f)_R$)-invariant contraction of any upper index with any lower index identically vanishes. In general, any tensor can be decomposed as a direct sum of terms, each being a traceless tensor or the product of one (lower-rank) traceless tensor times a certain number of invariant tensors $\delta^i_j$ (see for example Ref.~\cite{Tung:1985na}, chapter 13).}  transform as irreducible representations of the global symmetry group, and thus play an important role.
When representing traceless tensors, we will distinguish between upper (lower) indices of $SU(N_f)_L$, denoted as $n_L$ ($\bar n_L$), and upper (lower) indices of $SU(N_f)_R$, denoted as $n_R$ ($\bar n_R$). For each set of $n$ indices, one can consider an associated Young Tableau (YT), denoted as $\{n\}$ in the following, with $n$ boxes. We will use the compact notation $T^{\{n_L\};\{n_R\}}_{\{\bar{n}_L\};\{\bar{n}_R\}}$ to indicate traceless tensors of total upper rank $n = n_L +n_R$ and total lower rank $\bar n = \bar n_L +\bar n_R$. 
The corresponding operators will have $n_q = n +\Delta$ quarks and $n_{\bar q}=\bar n +\Delta$ antiquarks, where $\Delta$ is the number of quark and antiquark indices contracted in a flavor singlet.
Each traceless tensor satisfying the sum rule (\ref{eq:cons}) corresponds to a massless bound state.~\footnote{By massless bound state we mean a set of massless states with the same quantum numbers under the global symmetry ${\cal G}[N_f]$, whose degeneracy is accounted for by means of the index defined in Eq.~(\ref{eq:indexdef}).   Such states are indistinguishable using low-energy observables, since these depend only on the quantum numbers under~${\cal G}[N_f]$.}
For a given theory with $N_c$ colors and $N_f$ flavors, the set of all possible traceless tensors that satisfy Eq.~(\ref{eq:cons}) forms a tensor space denoted as ${\cal T}[N_f]$:
\bea
\text{Composite Operators} \sim \ T^{\{n_L\};\{n_R\}}_{\{\bar{n}_L\};\{\bar{n}_R\}}\ \in {\cal T}[N_f] \ .
\eea
For convenience, we will denote by ${\cal R}[N_f]$ the corresponding space of representations of ${\cal G}[N_f]$.
When no confusion arises, we will use the shorthand notation 
\begin{equation}
T^{\{n\}}_{\{\bar{n}\}} \equiv T^{\{n_L\};\{n_R\}}_{\{\bar{n}_L\};\{\bar{n}_R\}}\, ,
\end{equation}
where each of the symbols ${\{n\}}$ and ${\{\bar{n}\}}$ collectively denotes a pair of YTs with $n=n_L+n_R$ and $\bar{n}=\bar{n}_L+\bar{n}_R$.

Indices of $SU(N)$ can be raised (lowered) by means of the fully antisymmetric invariant tensor $\epsilon^{i_1 i_2\cdots i_{N}}$ ($\epsilon_{i_1 i_2\cdots i_{N}}$). To find the representation characterizing a given tensor one can proceed as follows~\cite{Tung:1985na}: for each kind of indices, starting from the (antiquark) YT ${\{\bar{n}\}}$ one finds the dual tableau $\widetilde{\{\bar n\}}$ obtained by replacing each column of length $\zeta$ with a column of length $N_f-\zeta$ (columns of $\widetilde{\{\bar n\}}$ must of course be arranged in reverse order compared to those in ${\{\bar{n}\}}$); then, the representation to be found is that given by the `combined' YT (CYT) obtained as the union of the columns of $\{n\}$ and $\widetilde{\{\bar n\}}$.~\footnote{For example, the adjoint representation of $SU(N_f)$ can be thought of as the traceless part of the tensor product of a fundamental and an anti-fundamental representation, which can be written as the tensor $T^{\{1\}}_{\{1\}}$. The CYT has two columns, one with $N_f-1$ boxes and one with a single box. According to the procedure outlined above, the CYT is obtained by first replacing $\{1\}$ from the lower index by a column of $N_f-1$ boxes, and then joining it with the box from the upper index.} 
\enlargethispage{-0.7cm}
For brevity, the CYT obtained from $\{n\}$ and $\{\bar{n}\}$ will be denoted as 
\bea
\{n;\bar{n}\}\equiv \{n\} + \widetilde{\{\bar{n}\}} \ , 
\eea
where the `sum' on the right-hand side means adding together the columns of $\{n\}$ and $\widetilde{\{\bar{n}\}}$ to make a new YT.

A massless bound state interpolated by a traceless tensor $T^{\{n_L\}, \{n_R\}}_{\{\bar{n}_L\}, \{\bar{n}_R\}}$ (i.e. by an operator corresponding to that tensor~\footnote{In the following, when we say that a tensor interpolates a bound state we mean that the bound state is interpolated by some local operator which corresponds to that tensor. }) transforms as an irrep $r$ of ${\cal G}[N_f]$ characterized by a pair of CYT and baryon number $b$ given by Eq.~(\ref{eq:cons}):
\bea
\label{irreps}
r = \left(\{n_L; \bar{n}_L\}, \{n_R; \bar{n}_R\}, b\right)\ .
\eea

\vspace{0.3cm}
\textbf{Example~2.1}. Consider $T^{\{3_a\};\{0\}}_{\{0\};\{0\}}$, where $\{ 3_a\}= \tiny\yng(1,1,1)\ $. Such tensor exists for $N_f\geq 3$ and satisfies the sum rule of Eq.~(\ref{eq:cons}) with $b=1$ for $N_c=3$. It interpolates a baryon transforming as $r=\left(\ {\tiny\yng(1,1,1)}\ ,\ s\ , 1\right)$, where $s$ stands for the singlet representation of $SU(N_f)$.

\subsection{Equivalent Tensors}
\label{sec:equivtensor}

In general, different traceless tensors of ${\cal T}(N_f)$ interpolate massless states in different representations of~${\cal G}[N_f]$, i.e. different massless states. There are some important exceptions however, which we call equivalent tensors.

First of all, tensors interpolating the same representation of ${\cal G}[N_f]$ can have YTs differing by the presence of one (or more) set of $N_f$ fully antisymmetrized indices, i.e. flavor singlets 
\bea
\epsilon_{i_1 i_2\cdots i_{N_f}} q^{i_1} q^{i_2}\cdots q^{i_{N_f}} \ . 
\eea
Tensors of this kind can still obey the sum rule (\ref{eq:cons}) with the same~$b$, and thus be equivalent under ${\cal G}[N_f]$, if such flavor singlets are located in different YTs. 

\vspace{0.3cm}
\textbf{Example~2.2}. Consider $T^{\{N_f\};\{0\}}_{\{1\};\{0\}}$ and $T^{\{ 0\};\{N_f\}}_{\{1\};\{0\}}$, for $bN_c = N_f-1$, where $\{ N_f\}$ and $\{1\}$ are respectively a YT with a single column of length $N_f$ and a YT with one box. Both these tensors transform as an anti-fundamental of $SU(N_f)_L$, a singlet representation of $SU(N_f)_R$, and have baryon number $b=(N_f-1)/N_c$.
\vspace{0.3cm}

There is another, less trivial, way in which equivalent tensors can arise. Notice that two different pairs of upper and lower YTs (for either left or right indices) can correspond to the same representation of $SU(N_f)$.
Two tensors are equivalent under ${\cal G}[N_f]$, i.e. correspond to the same representation, if their CYTs for both left and right indices are identical and their baryon number is the same. 

\vspace{0.3cm}
\textbf{Example~2.3}.
Consider $T^{\{0\};\{n\}}_{\{0\};\{m\}}$ and $T^{\{0\};\{N_f-m\}}_{\{0\};\{N_f-n\}}$ with $n, m < N_f$ and $bN_c=n-m$, where all the indices are fully antisymmetrized. These two equivalent tensors transform trivially under $SU(N_f)_L$ and as an irrep of $SU(N_f)_R$ characterized by a CYT with two columns, one with $n$ boxes and one with $N_f-m$ boxes. They have the same baryon number $b=(n-m)/N_c$. The case with $m=0$ gives an example of a tensor with lower (antiquark) indices interpolating a baryonic bound state:  $T^{\{0\};\{N_f\}}_{\{0\};\{N_f-n\}}$ is equivalent to the purely baryonic operator $T^{\{0\};\{n\}}_{\{0\};\{0\}}$.
\vspace{0.3cm}

\begin{figure}[t]
\centering
\includegraphics[scale=0.26]{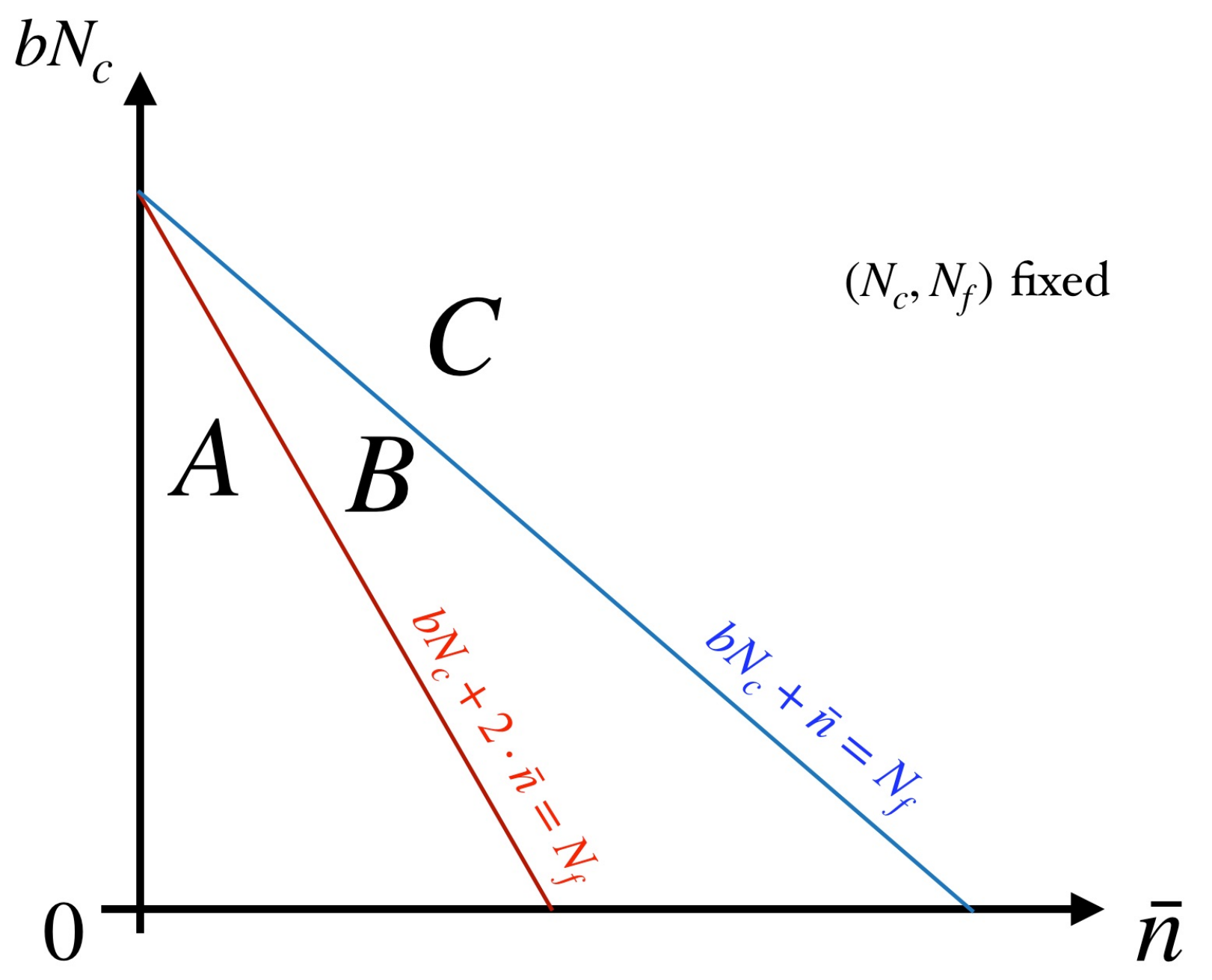}
\caption{Different classes of tensors that interpolate massless bound states in a confining description of QCD-like theory with $N_c$ colors and $N_f$ flavors, under the assumption of unbroken chiral symmetry.  Tensors are defined in terms of their baryon number $b$ (c.f. Eq.~(\ref{eq:cons})) and antiquark number $\bar n = \bar n_L +\bar n_R$.}
\label{fig:spectrum}
\end{figure}

As it will become clear in the following, understanding the properties of equivalent tensors helps us to understand the validity of $N_f$-independence and of the superalgebra approach. 
It is useful to classify traceless tensors into three classes according to their baryon number $b$ and their ranks $\bar n = \bar n_L+\bar n_R$ and $n=n_L+n_R$, as depicted in Fig.~\ref{fig:spectrum}. 
%
%
For fixed $N_c$ and $N_f$, the three classes are defined as follows:
\begin{itemize}
\item Class $A$: tensors with $n+\bar n = 2\bar n + bN_c < N_f$. 
\item Class $B$: tensors with $n = \bar n + bN_c \leq N_f \leq 2\bar n + bN_c  = n + \bar n$; when $n=N_f$ a flavor singlet can appear in the traceless part (i.e. $\epsilon_{i_1 i_2\cdots i_{N_f}} q^{i_1} q^{i_2}\cdots q^{i_{N_f}}$), otherwise they contain no flavor singlets; since $\bar n < n$, given that $b>0$, no flavor singlets made of antiquark indices can appear.
\item Class $C$: tensors with $n = \bar n + bN_c > N_f$.
\end{itemize}
The following result holds true:
\\[0.25cm] 
\textbf{Lemma 1 (Absence of Equivalent Tensors).} \  \textit{Any two class-A tensors cannot transform in the same representations under $\mathcal{G}[N_f]$, i.e. they are not equivalent. Any two terms in their tensor decomposition under $\mathcal{G}[N_f,1]$  (cf. Eq.~(\ref{def:G_prime})) which have non-zero $U(1)_{H_1}$ charge are not equivalent either.
\\[0.4cm]
}
The proof is given in Appendix~\ref{app:eq_tensor}.

\section{Anomaly Matching and Persistent Mass Conditions}
\label{sec:overview}

The main purpose of this section is to give a review on AMC and PMC (see also~\cite{Weinberg:1996kr}). Some important properties of the PMC overlooked in the previous literature will be also discussed.
  
In order to write the AMC and PMC, we first introduce the index $\ell\!\left(r\right)$ to denote the multiplicity of the (massless) bound states in the irrep $r$, defined as the number of times these appear in the spectrum with helicity $+1/2$ minus the number of times they appear with helicity~$-1/2$~\cite{tHooft:1979rat}:
\begin{equation}
\label{eq:indexdef}
\ell\!\left(r\right)\equiv {\cal N}\!\left(r\right)_{+1/2} - {\cal N}\!\left(r\right)_{-1/2}\ .
\end{equation}
Since ${\cal N}\!\left(r\right)_{\pm 1/2}$ are positive integers, the indices $\ell$ are also integers. The invariance of the theory under CPT implies that
\bea
{\cal N}\!\left(r\right)_{-1/2}= {\cal N}\!\left(r^*\right)_{+1/2}\, .
\eea
Here $r^*$ is the complex conjugate of $r$,  in particular it has opposite baryon number.
This property allows us to consider only  bound states with positive baryon number without loss of generality: the multiplicity of a massless bound state with negative baryon number is equal and opposite to the multiplicity of its antiparticle, whose baryon number is positive. 

Parity is unbroken in vectorlike gauge theories with massive fermions and $\theta=0$~\cite{Vafa:1984xg}. Continuity implies that the parity-preserving vacuum is a global minimum also in the limit of vanishing quark masses, although degenerate parity-breaking minima might appear in principle.  If the vacuum preserves parity, then it follows, for any $r = \left(r_L, r_R, b\right)$, that
\bea
{\cal N}\!\left(r\right)_{-1/2} = {\cal N}\!\left(r_P\right)_{+1/2}\, ,
\eea
where $r_P = \left(r_R, r_L, b\right)$ is the parity conjugate of $r$.
Notice however that our results do not rely on the assumption of unbroken parity.

\subsection{Anomaly Matching Conditions}
\label{sec:amc}

There are four 't Hooft anomalies that must be matched in the IR by composite massless fermions: those of $[SU(N_f)_{L}]^2 U(1)_B$, $[SU(N_f)_{L}]^3$, $[SU(N_f)_{R}]^2 U(1)_B$ and $[SU(N_f)_{R}]^3$.~\footnote{If parity is spontaneously broken by the vacuum, the spectrum of massless bound states can be chiral under $U(1)_B$, and matching the $[U(1)_B]^3$ anomaly implies one extra constraint (if parity is unbroken, the matching is automatic). In the following, we will not make use of this additional anomaly matching equation.}
The AMC of $[SU(N_f)_{L}]^2 U(1)_B$ and $[SU(N_f)_{L}]^3$ can be written explicitly as
\begin{equation}
\label{eq:am}
\sum_{r\in {\cal R}[N_f]} \,\ell\!\left(r \right) A_i\!\left(r\right) =N_c \, A_{i}\!\left(r_{q_L}\right),
\end{equation}
where $A_i(r)$ is the anomaly contribution of the representation $r$ of ${\cal G}[N_f]$, and where $i=2$ for $[SU(N_f)_{L}]^2 U(1)_B$ and $i=3$ for $[SU(N_f)_{L}]^3$. The left-hand side of this equation encodes the anomaly of the bound states, expressed as a sum over the contributions of all possible representations $r$ in the spectrum multiplied by their index $\ell$;  the right-hand side encodes instead the anomaly of the left-handed quarks $q_L$, which transform in the representation $r_{q_L} = (\tiny\yng(1),s,1/N_c)$. Here and in the following we assume $N_f>2$. For $N_f=2$, since all the representations of $SU(2)$ are real or pseudo-real, there is only one AMC, that of $[SU(2)_{L}]^2 U(1)_B$, which has integer solutions~\cite{tHooft:1979rat,Cohen:1981iz}.

The anomaly of a representation $r = \left(r_L, r_R, b\right)$ can be computed as
\begin{equation}
A_i(r) = d\!\left(r_R\right)  D_{i}\!\left(r_L\right)\, , 
\end {equation}
where $d\left(r_R\right)$ is the dimension of $r_R$, and $D_{i=2,3}\!\left(r_L\right)$ are the anomaly traces of respectively $[SU(N_f)_{L}]^2 U(1)_B$ and $[SU(N_f)_{L}]^3$, see Appendix~\ref{app:trace} for their definitions. Both quantities depend on $N_f$ explicitly. For the quarks, in particular, $A_i(r_{q_L}) = D_i(\tiny\yng(1))$. 
By conveniently defining the following ratios of anomaly traces
\bea
\bar D_{i}\!\left( r_L \right) \equiv \frac{D_{i}\!\left(r_L\right)}{D_{i}\!\left(\tiny\yng(1)\right)}\ , \qquad i=2,3
\label{eq:ano_const}
\eea
the AMC can be recast into the form 
\bea
\sum_{r\in {\cal R}[N_f]} \,\ell\!\left(r \right)
d\!\left(r_R\right)  \bar D_{i}\!\left(r_L\right)=N_c \, \ .
\label{eq:am_2}
\eea
The AMC for $[SU(N_f)_{R}]^2 U(1)_B$ and $[SU(N_f)_{R}]^3$ can be derived in a similar way. One thus has a system of four linear, non-homogeneous equations for the variables $\ell$, that will be denoted as AMC$[N_f]$ in the following to stress its dependence on the number of flavors~$N_f$.

If $N_c$ is even, any color-singlet bound state is necessarily a boson. This directly implies that 't~Hooft anomalies cannot be matched unless chiral symmetry is spontaneously broken and massless Nambu-Goldstone bosons are present in the spectrum. In theories with $N_c$ odd, proving chiral symmetry breaking is not equally easy. 
It is possible to show that AMC[$N_f$] do not have integer solutions when $N_f$ is an integral multiple of a prime factor $p>1$ of $N_c$, see Ref.~\cite{Ciambriello:2024xzd}. For such special values of $N_f$ chiral symmetry breaking is therefore proven. A first application of this argument was made by Preskill and Weinberg in Ref.~\cite{Preskill:1981sr} in the context of three-flavor QCD and for the case of baryons with $b=1$, see also~\cite{Weinberg:1996kr}.
In the following, we will discuss the case of theories with $N_c$ odd and $N_f$ generic.

\subsection{Persistent Mass Conditions}
\label{sec:pmc}


In its original formulation, the Persistent Mass Condition asserts that any bound state containing massive constituents must also be massive~\cite{Preskill:1981sr}. It is however possible to give the following more precise statement that does not make reference to internal constituents of the bound states and relies instead only on their representation under the chiral symmetry group.~\footnote{Although intuitive, the notion of internal constituents can be made precise only in the framework of a given phenomenological model for the bound states and, as such, it is not convenient.}

When one flavor is given a mass, each massless fermionic bound state transforming as an irrep of ${\cal G}[N_f]$ breaks up into irreps of the unbroken flavor group $\mathcal{G}[N_f,1]$. The Persistent Mass Condition requires that, after such decomposition, the bound states with nonzero $U(1)_{H_1}$ charge must be massive, i.e. come in vectorlike pairs with vanishing index. More in general, when $i$ flavors are given unequal masses, all the bound states that are charged under $U(1)_{H_1} \times \cdots \times U(1)_{H_i} \subset {\cal G}[N_f,i]$ must be massive, i.e. have vanishing index.

As formulated above, the Persistent Mass Condition can be proven by using the arguments articulated by Vafa and Witten in Ref.~\cite{Vafa:1983tf}. Let us analyze the case of one massive flavor, the generalization to additional massive flavors being straightforward. We consider a regularized version of our theory with finite cutoff $\Lambda$.~\footnote{Our argument, as well as those of~\cite{Vafa:1983tf}, is formulated for a theory on the continuum. One might want to resort to lattice regularization for a fully rigorous definition of the theory. Notice that the assumptions on which the Vafa-Witten theorem relies are fulfilled only on specific lattice realizations, like that given by staggered fermions. See for example~\cite{Aloisio:2000rb,Azcoiti:2010ns,Giordano:2023spj}.}
Let $m$ be the bare mass of the massive flavor and $\epsilon$ that of all the remaining flavors. At the end we will take the limit $\epsilon\to 0$ and then remove the cutoff.
We stress that $m$ and $\epsilon$ are bare and not renormalized masses. Consider a local operator $T(x)$ that interpolates bound states with $U(1)_{H_1}$ charge $H_1 \not = 0$.
The results of Ref.~\cite{Vafa:1983tf} imply that the two-point function of $T(x)$ satisfies the bound
\bea
| \langle T^\dagger(x) T(y)\rangle | \lesssim C  \, e^{-(Q m+q\epsilon)|x-y|}
\eea
where $Q = |H_1|$, $q=|bN_c-H_1|$ and $C$ is a prefactor that does not depend on $x$ and $y$. Such exponential fall-off means that any bound state excited by $T(x)$ has mass $M \geq Q m+q\epsilon$, where of course $M$ depends on~$\Lambda$. If we now let $\epsilon\to 0$ we obtain the lower bound $M \geq |H_1| m >0$. This result holds in the regularized theory with finite cutoff. It states that for any finite value of $\Lambda$, the spectrum of the bound states excited by $T(x)$ must be vectorlike, i.e. the corresponding index is~0. Since the theory is defined as the limit for $\Lambda\to \infty$ of the regularized one, this means that the index will vanish also after the cutoff is removed.~\footnote{If one uses the renormalization group to resum the leading logs, one finds that the bare mass $m$ goes to zero if the limit $\Lambda\to \infty$ is taken while keeping the renormalized mass fixed and finite. We thank M.~Bochicchio for pointing this to us. While this seems to nullify the bound on $M$, the spectrum of bound states excited by $T(x)$ remains vectorlike and this is sufficient to prove that the index is zero after removing the cutoff.} This proves the PMC.

Notice that it was crucial to assume $H_1\not =0$ for the operator $T(x)$. No firm statement can be made on bound states with $H_1=0$, even though they can be interpolated by operators featuring massive quarks and antiquarks. In fact, since $H_1=0$, there will necessarily exist at least one operator \emph{without} heavy flavors that can interpolate those states. This also means that decoupling cannot be used to forbid massless states with $H_1=0$.  Indeed, decoupling is just the statement that one can reproduce physical results at low energy by means of an effective theory defined in terms of degrees of freedom with $H_1=0$.
To the best of our knowledge, the proof of PMC and the importance of the condition $H_1\not =0$ were not explicitly discussed in the previous literature. Going beyond a naive formulation of PMC is especially required when dealing with exotic states.

On the mathematical level, the Persistent Mass Condition implies a series of equations which must be satisfied by the indices $\ell$.
To discuss this, it is useful to introduce the following notation: we will denote by
\begin{itemize}
 \item ${\cal R}_0[N_f,i]$ the space of irreps of ${\cal G}[N_f,i]$ with all vanishing charges ($H_k=0$ with $k=1,\dots, i$);
\item $\hat {\cal R}[N_f,i]$ the space of irreps of ${\cal G}[N_f,i]$ with $H_i\not = 0$ and $H_k=0$ for $k=1,\dots, i-1$.
\end{itemize}
In the case of one massive flavor, the index of any irrep $\hat r_1$ of ${\cal G}[N_f,1]$ with $H_1\not =0$ must vanish:
\begin{equation}
\label{eq:pmc}
0 =  \ell\!\left( \hat r_1 \right) = \sum_{r\in {\cal R}[N_f]}\, k\!\left(r \to \hat r_1 \right)  \ell\!\left(r\right) \, , \qquad \forall \ \hat r_1\in \hat{\cal R}[N_f,1]\, .
\end{equation}
The coefficient $k(r \to \hat r_1 )$ equals the number of times $\hat r_1$ appears in the decomposition of~$r$.  Let $\hat r_1 = (\{m_L ; \bar{m}_L \}, \{m_R; \bar{m}_R \}, H_1, b)$. It is easy to see that
\bea
H_1 = bN_c - (m_L+m_R) + (\bar{m}_L+\bar{m}_R)\, , 
\label{eq:H_1}
\eea
while the baryon number $b$ is the same as that of the parent irrep $r$.
The system of all PMC (one equation for each irrep $\hat r_1\in \hat {\cal R}[N_f,1]$) that are obtained by giving one of the $N_f$ flavors a mass will be denoted as $\text{PMC}[N_f,1]$ in the following. 
Notice that $\text{PMC}[N_f,1]$ depend implicitly on $N_f$, i.e. in general they have different expressions in confining theories with different number of flavors. This point will be elaborated further in Section.~\ref{sec:proof_Nf_independence}. 

Additional PMC arise when more than one flavor is given a mass. For example, when a second flavor becomes massive, those states which had $H_1=0$ in the first step will now break up into multiplets of the unbroken flavor group ${\cal G}[N_f,2]$, cf. Eq.~(\ref{def:G_prime}).
After this second step of decomposition, the states which are charged nontrivially under $U(1)_{H_2}$ are subject to $\text{PMC}[N_f,2]$, i.e. equations of the form
\begin{equation}
\label{eq:pmc_Nf_2}
0 =  \ell\!\left( \hat r_2 \right) = \sum_{r_1\in {\cal R}_0[N_f,1]}\, k\!\left(r_1 \to \hat r_2 \right)  \ell\!\left(r_1\right)\ , \qquad \forall \ \hat r_2\in \hat{\cal R}[N_f,2]\, ,
\end{equation}
where the sum runs over all irreps $r_1$ of ${\cal G}[N_f,1]$ with $H_1=0$. Let $\hat r_2 = (\hat r_{2L}, \hat r_{2R}, H_1=0, H_2, b)$, with $\hat r_{2L}=\{m^\prime_L ; \bar{m}^\prime_L \}$, $\hat r_{2R}=\{m^\prime_R; \bar{m}^\prime_R \}$; then it follows that
\bea
H_2 = bN_c - (m^\prime_L+m^\prime_R) + (\bar{m}^\prime_L+\bar{m}^\prime_R). 
\label{eq:H_2}
\eea
Notice that $\text{PMC}[N_f,2]$ also depend implicitly on $N_f$.
More in general, in a theory with $N_f$ flavors there exist $N_f-2$ sets of PMC, denoted by $\text{PMC}[N_f,i]$ with $1 \leq i\leq N_f-2$, obtained by considering all possible numbers of massive flavors and giving them unequal masses. Giving the same mass to different flavors does not lead to new equations, as shown in Appendix~\ref{app:PMCsamemasses}. Since there are no PMC in a theory with 2 flavors, the largest number of flavors that can be made massive to obtain new equations is $N_f-2$.~\footnote{If $N_f = 2$ and one flavor is given a mass, the unbroken group $G[2, 1] = U(1)_H \times U(1)_B$ is vectorlike. If parity is unbroken in the vacuum, all the indices of irreps of $G[2, 1]$ vanish automatically. On the other hand, if parity is broken by the vacuum, then new non-trivial PMC equations arise. In the following, we will not make use of such additional PMC equations.}
The collection of all the PMC in a theory with $N_f$ flavors will be denoted by $\text{PMC}[N_f]$ in the following.
To the best of our knowledge, previous works only considered PMC with one massive flavor~\cite{tHooft:1979rat,Frishman:1980dq, Schwimmer:1981yy, Farrar:1980sn, Takeshita:1981sx, Kaul:1981fd, Cohen:1981iz,Bars:1981nh}; these are denoted as PMC$[N_f,1]$ in this paper. On the other hand, PMC with more than one massive flavor play an important role in the proof of chiral symmetry breaking (see Ref.~\cite{Ciambriello:2024xzd}).

\begin{figure}[t]
\centering
\includegraphics[scale=0.4]{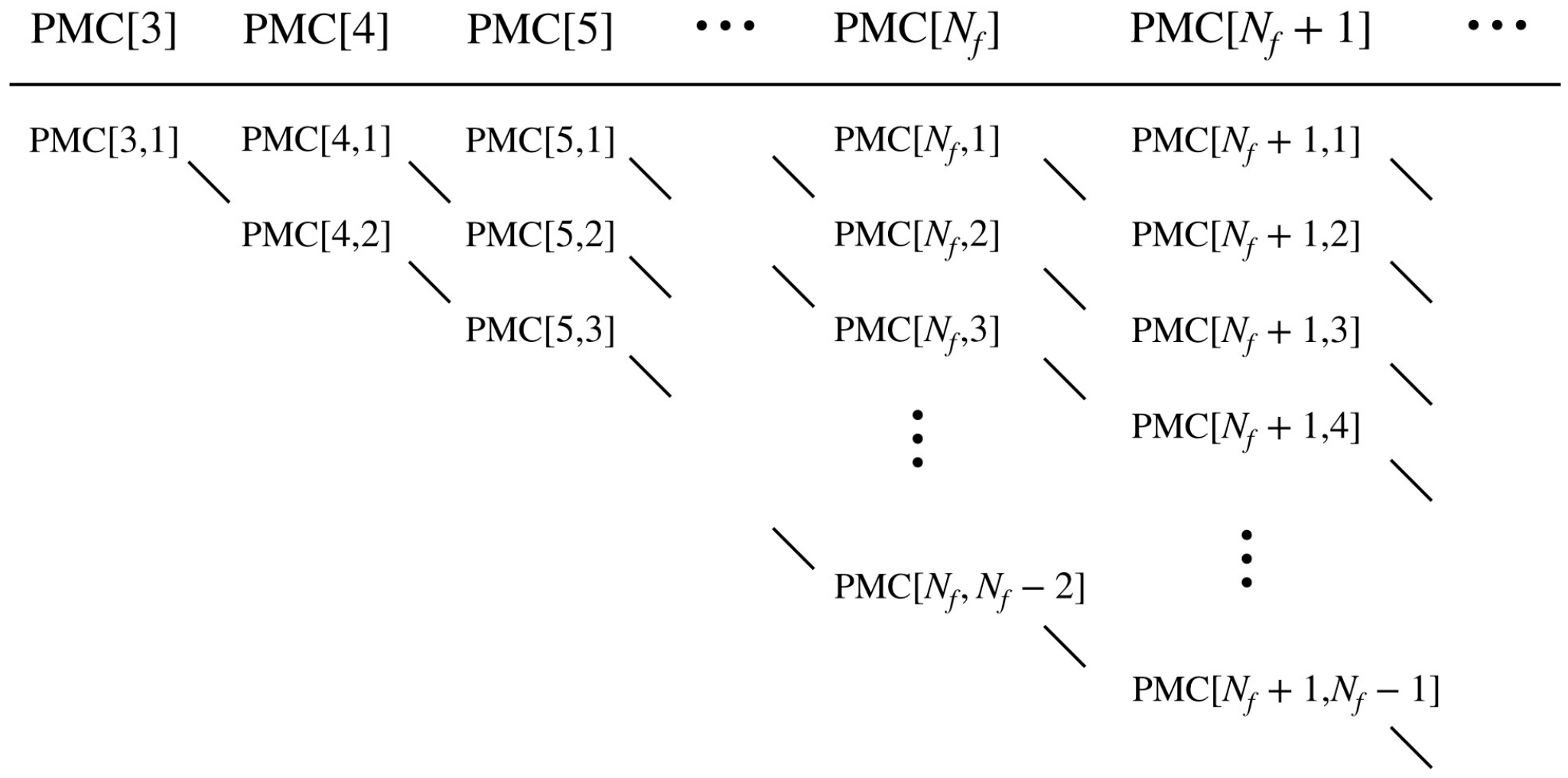}
\caption{General structure of persistent mass conditions $\text{PMC}[N_f]$ for different values of $N_f$. The set of PMC obtained by giving different masses to $i$ flavors, with $1 \leq i\leq N_f-2$, is denoted as $\text{PMC}[N_f,i]$. Sets of PMC equations connected by a line have the same form.}
\label{fig:pmc}
\end{figure}

It is useful to notice that $\text{PMC}[N_f,2]$ are related to $\text{PMC}[N_f-1,1]$.  Indeed, each of the bound states that remain massless after giving one flavor a mass is in one-to-one correspondence with a massless bound state of a theory with $N_f-1$ flavors. As a consequence, $\text{PMC}[N_f,2]$ have the same form as $\text{PMC}[N_f-1,1]$. Indeed, the latter are written as
\begin{equation}
\label{eq:pmcp}
0 =  \ell\!\left( \hat r'_1\right) = \sum_{r'\in {\cal R}[N_f-1]}\, k\!\left(r' \to \hat r'_1 \right)  \ell\!\left(r'\right) \, , \qquad \forall \ \hat r'_1\in \hat{\cal R}[N_f-1,1]\, ,
\end{equation}
which is identical to Eq.~(\ref{eq:pmc_Nf_2}) provided one identifies $r'$ with $r_1$ and sets $\ell\!\left(r'\right) = \ell\!\left(r_1\right)$. The identification is possible because ${\cal R}_0[N_f,1] = {\cal R}[N_f-1]$, i.e. any irrep of ${\cal G}[N_f,1]$ with $H_1=0$ can be seen as an irrep of ${\cal G}[N_f-1]$.
More in general, it is simple to see that $\text{PMC}[N_f,i]$ have the same form as $\text{PMC}[N_f-1,i-1]$, for $1< i \leq N_f-2$. The general structure of PMC when $N_f$ is varied is therefore the one sketched in Fig.~\ref{fig:pmc}.

\section{Proof of Chiral Symmetry Breaking from $N_f$-independence}
\label{sec:Nf-independence}

The $\text{AMC}[N_f]$ and $\text{PMC}[N_f]$ give a linear system of equations on the indices $\ell$. As first suggested by 't Hooft~\cite{tHooft:1979rat}, it is useful to consider such a system and its solutions as functions of $N_f$. This is independent of whether the theory with $N_f$ flavors is in the confining regime or not; one can indeed consider Eq.~(\ref{eq:am}) as a rule to construct the $\text{AMC}[N_f]$ for any~$N_f$, and similarly Eqs.~(\ref{eq:pmc}) and~(\ref{eq:pmc_Nf_2}) as rules to construct the $\text{PMC}[N_f]$.
Under particular assumptions that will be discussed in detail, the following property then holds and is key to prove chiral symmetry breaking:
\\[0.25cm] \textbf{$N_f$-independence}: \  \textit{Any finite set of real indices $\{\ell\}$ that solves the system of AMC and PMC for a theory with $N_f$ flavors is also a solution of the same equations for any $N'_f \geq N_f$.}
\\[0.4cm]
In order to properly state $N_f$-independence one should better qualify what it exactly means that the same set of indices $\{\ell\}$ is a solution of AMC and PMC for different numbers of flavors. As we will show, one needs to establish a map between irreps of the flavor groups ${\cal G}[N_f]$ and ${\cal G}[N'_f]$, i.e. a map between states of different theories. For the moment we will content ourselves with the naive definition given above and explore its consequences.

The validity of $N_f$-independence implies that chiral symmetry must be broken in the theory with $N_f$ flavors. 
There are at least two ways to see this. The first makes use of the following simple argument: let us suppose that for $N_f$ flavors and $N_c$ colors the theory is in the confining regime while chiral symmetry is not spontaneously broken by the vacuum. This requires that AMC$[N_f]$ have integer solutions. By $N_f$-independence, this in turn implies that AMC$[N_f']$ must also have integer solutions for any $N_f'\geq N_f$. However, we know that AMC$[p\, m]$ do not admit integer solutions for any positive integer $m$, where $p$ is any prime factor of $N_c$. This contradicts the initial assumption and thus proves (as long as $N_f$-independence is valid) that chiral symmetry is spontaneously broken in the theory with $N_f$ flavors. Notice that it is not necessary to assume that the theory with $N_c$ colors and $N_f' > N_f$ flavors is in the confining regime.

A second way to see that chiral symmetry breaking is implied by $N_f$-independence is by means of the following argument.
The anomaly coefficients $A_i(r)$ have polynomial dependence on $N_f$. Therefore, as long as $N_f$-independence holds, the AMC can be recast, for any $N'_f\geq N_f$, as
\bea
\label{eq:polynom}
a_h\!\left(\{\ell\}\right) (N'_f)^{h} + a_{h-1}\!\left(\{\ell\}\right) (N'_f)^{h-1}+\cdots+ a_1\!\left(\{\ell\}\right) N'_f+ a_0\!\left(\{\ell\}\right) =0\ ,
\eea
where the coefficients $a_i\!\left(\{\ell\}\right)$ depend on the indices $\ell$. In other words, the polynomial in $N'_f$ of Eq.~(\ref{eq:polynom}) generates the AMC for any $N'_f\geq N_f$. Tensors with $n$ upper and $\bar n$ lower indices contribute to the AMC with terms of order $n+\bar n-1$ or smaller in $N_f$. The highest degree of the polynomial, $h$, is thus finite if the spectrum of representations, hence of interpolating tensor operators, is finite:
\be
h = n_{max} + \bar n_{max} -1=bN_c + 2 \bar n_{max}-1\ . 
\ee
$N_f$-independence implies that Eq.~(\ref{eq:polynom}) has a solution for every integer $N_f' \geq N_f$ once its coefficients are evaluated on the set of indices $\{\ell\}$ which solves $\text{AMC}[N_f]\cup\text{PMC}[N_f]$. But a polynomial can have a number of solutions bigger than its degree only if it vanishes identically. This implies that its coefficients must individually vanish:
\bea
\label{eq:polynom_2}
a_i\!\left(\{\ell\}\right)=0, \quad\text{for all}\ \ 0\leq i\leq h\ .
\eea
As pointed out in Ref.~\cite{Farrar:1980sn} by Farrar, the equation $a_0\!\left(\{\ell\}\right)=0$ does not admit integral solutions.
Hence, no integral solution of the $\text{AMC}[N_f]$ exists, and this means that the assumption of unbroken chiral symmetry is not consistent: chiral symmetry must be spontaneously broken in the theory with $N_f$ flavors.

\subsection{Proof of $N_f$-independence}\label{sec:proof_Nf_independence}

In the past literature $N_f$-independence was assumed to hold based on empirical arguments. Given its centrality to prove chiral symmetry breaking, we want to derive a rigorous proof and spell out explicitly the assumptions which underlie its validity. Following Refs.~\cite{Frishman:1980dq,Takeshita:1981sx,Kaul:1981fd}, we can first check whether $\text{PMC}[N^\prime_f]$ can be solved, for any $N^\prime_f\geq N_f$, by the same indices that solve $\text{PMC}[N_f]$, and then check whether a solution of $\text{AMC}[N_f]$ is also a solution of $\text{AMC}[N^\prime_f]$, up to equations implied by $\text{PMC}[N^\prime_f]$. If the solution of $\text{AMC}[N_f]\cup\text{PMC}[N_f]$  is also a solution of $\text{AMC}[N^\prime_f]\cup\text{PMC}[N^\prime_f]$, then $N_f$-independence follows. We find that this is possible only for a spectrum of massless bound states that are interpolated by class-A tensors.

\subsubsection{PMC$\,[N_f]$ are $N_f$-dependent in general}

In order to see if a real solution of $\text{PMC}[N_f]$ is also a real solution of $\text{PMC}[N^\prime_f]$ for $N^\prime_f > N_f$ we start by considering $N^\prime_f =N_f+1$ and then proceed iteratively. The relation between $\text{PMC}[N_f]$ and $\text{PMC}[N_f+1]$ is clarified by recalling that $\text{PMC}[N_f+1,i]$ have the same form as $\text{PMC}[N_f,i-1]$, for $1< i \leq N_f-1$, see Section~\ref{sec:pmc}. This means that  $\text{PMC}[N_f+1]$ contain $\text{PMC}[N_f]$ plus the additional set of equations $\text{PMC}[N_f+1,1]$. It is thus crucial to understand whether the latter are truly linearly-independent equations. We will now show that $\text{PMC}[N_f+1,1]$ are identical to $\text{PMC}[N_f,1]$ under specific conditions.

The form of $\text{PMC}[N_f,1]$ is in general different in theories with different numbers of flavors because of the following three reasons:
\begin{enumerate}
\item New tensor structures arise for $N_f^\prime>N_f$. In general, such new tensor structures imply new irreps in the theory with $N_f^\prime$ flavors. For example, tensors with $bN_c$ fully antisymmetrized upper indices exist only for $N_f \geq bN_c$ but not for $N_f < bN_c$. Indeed, the $bN_c$ boxes can only be arranged such that the YT has at most $N_f$ rows. Clearly, there are more irreps allowed for $N_f \geq bN_c$ than for $N_f < bN_c$. This point was emphasized in Ref.~\cite{Cohen:1981iz} for a spectrum of massless baryons with $b=1$.
\item  Different tensors of ${\cal G}[N_f]$ can become equivalent or cease to be so when the number of flavors is increased. In other words, two different tensors can interpolate the same irrep of ${\cal G}[N'_f]$ but different irreps of ${\cal G}[N_f]$, or viceversa.
\item Different tensors of ${\cal G}[N_f,1]$ can become equivalent or cease to be so when the number of flavors is increased. Accordingly, a single PMC equation splits into different PMC equations as tensors become inequivalent, while different PMC equations collapse into a single equation when tensors become equivalent. 
\end{enumerate}
The last point is best clarified by looking at an explicit example.

\vspace{0.3cm}
\textbf{Example 4.1}. Let us consider a theory with $N_c=3$ and the following two tensors
\begin{equation}
T_1 = T^{\{1\}; \{3_s\}}_{\{1\}; \{0\}}, \qquad  T_2 = T^{\{3_a\}; \{2_s\}}_{\{2_a\}; \{0\}}\; ,
\end{equation}
where $\{3_s\} = \tiny\yng(3)\,$, $\{3_a\} = \tiny\yng(1,1,1)\,$, $\{2_s\} = \tiny\yng(2)\,$, $\{2_a\} = \tiny\yng(1,1)\,$. 
The decomposition of $T_1$ and $T_2$ in  tensors of~${\cal G}[N_f,1]$ contains the following two terms, both with $H_1=3$:
\begin{equation}
  \begin{split}
T_1 &= T_1' + \cdots \qquad , \quad  T_1' = T^{\{1\}; \{0\}}_{\{1\}; \{0\}} \\[0.2cm]
T_2 &= T_2' + \cdots \qquad , \quad T_2' = T^{\{2_a\}; \{0\}}_{\{2_a\}; \{0\}} 
 \end{split}
\end{equation}
where the dots indicate other terms in the decomposition.
For $N_f=4$, $T_1$ and $T_2$ transform respectively as the irreps
\begin{equation}
r_1 = \left(\ {\tiny\yng(2,1,1)}\ , \ {\tiny\yng(3)}\ ,\ 1\right)\, ,\quad\quad  r_2 = \left(\ {\tiny\yng(2,2,1)}\ , \ {\tiny\yng(2)}\ ,\ 1\right)
\end{equation}
of ${\cal G}[4]$ and are thus inequivalent. The tensors $T_1'$ and $T_2'$, on the other hand, both transform as the representation $\left(\ {\tiny\yng(2,1)}\ , s, \ 3,\ 1\right)$  of ${\cal G}[4,1]$ and are thus equivalent. The PMC$[4,1]$ equation corresponding to that representation reads 
\bea
\label{eq:pmc_example4.1}
0=\ell (r_1)+\ell (r_2) + \cdots 
\eea
where the dots indicate contributions from operators other than $T_1$ and $T_2$.
For $N_f=5$, $T_1$ and $T_2$ transform as the following irreps
\begin{equation}
r_1 = \left(\ {\tiny\yng(2,1,1,1)}\ , \ {\tiny\yng(3)}\ ,\ 1\right)\, ,\quad\quad  r_2 = \left(\ {\tiny\yng(2,2,2)}\ , \ {\tiny\yng(2)}\ ,\ 1\right)
\end{equation}
of ${\cal G}[5]$ while $T_1'$ and $T_2'$ transform respectively as
\begin{equation}
\left(\ {\tiny\yng(2,1,1)}\ , \ s, \ 3,\ 1\right)\qquad \text{and}\qquad  \left(\ {\tiny\yng(2,2)}\ , \ s, \ 3,\ 1\right)\, .
\end{equation}
The latter are different irreps of ${\cal G}[5,1]$, which means that $T_1'$ and $T_2'$ are inequivalent for $N_f=5$.
As a result, Eq.~(\ref{eq:pmc_example4.1}) splits into two equations of PMC$[5,1]$:
\bea
\label{eq:pmc_example4.1_2}
0=\ell (r_1)+\cdots \ ,\quad\quad 0=\ell (r_2) +\cdots\ .
\eea
In a similar fashion, one can construct examples where, as $N_f$ is increased, different PMC equations collapse into a single PMC equation because tensors of ${\cal G}[N_f,1]$ become equivalent.

\subsubsection{Condition for $N_f$-independence of PMC$\,[N_f,1]$}

We have shown that the form of PMC$[N_f,1]$ is $N_f$-dependent in general, which implies that $\text{PMC}[N_f+1,1]$ are linearly independent with respect to $\text{PMC}[N_f,1]$. This is in turn an obstruction to the $N_f$-independence of the solution of PMC. However, $N_f$-independence can be still obtained if one imposes suitable restrictions on the spectrum of possible massless states. 
In order to determine such restrictions, we introduce the notion of uplifting tensors.

\begin{figure}
	\centering
        \includegraphics[scale=0.26]{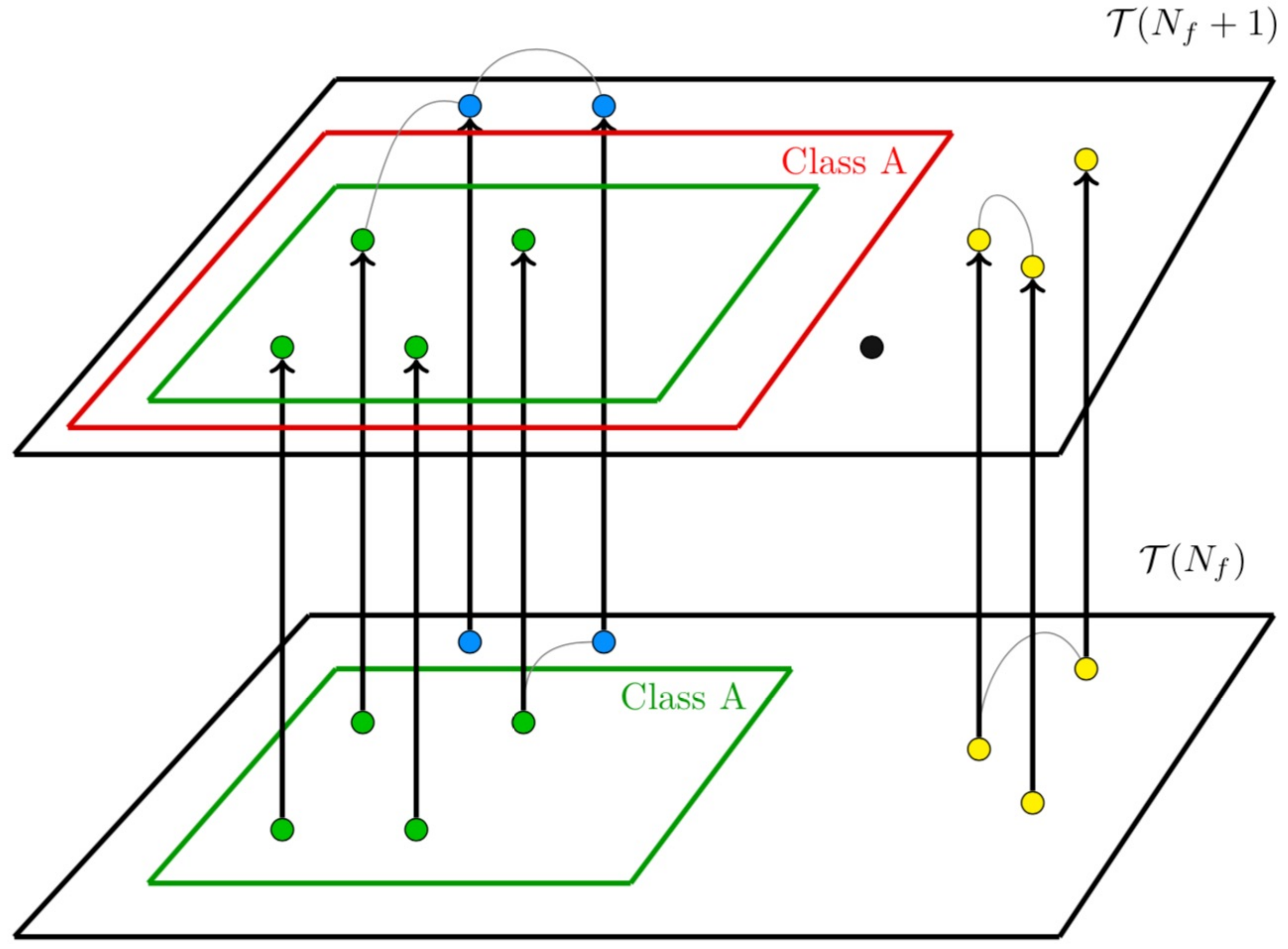}
	\caption{This cartoon tries to capture some of the properties of tensor uplifting. 
          The two planes represent the space of tensors ${\cal T}(N_f)$ and ${\cal T}(N_f+1)$, the dots (whose color corresponds to a particular property) being the individual tensors. The uplifting map is represented by the upward black arrows, while the curved gray lines represent the equivalence relation between equivalent tensors. The green box in the lower plane represents the subset of class-A tensors in ${\cal T}(N_f)$; its image through uplifting is represented by the green box in ${\cal T}(N_f+1)$. The latter is a proper subset of the set of class-A tensors in ${\cal T}(N_f+1)$, pictured by the red box. In this cartoon one can notice some of the properties explained in the text, in particular: \textit{i)} two class-A tensors cannot be equivalent; \textit{ii)} outside class-A tensors, the equivalence classes are not respected under uplifting (see the yellow dots); \textit{iii)} the image of ${\cal T}(N_f)$ through uplifting does not coincide with ${\cal T}(N_f+1)$ (the black dot is not the uplifted of any tensor in ${\cal T}(N_f)$). }
	\label{fig:cartoon_uplifting}
\end{figure}

We define \textit{uplifting} as a map between traceless tensors of ${\cal T}[N_f]$ and ${\cal T}[N_f+1]$. Given any tensor $T\in{\cal T}[N_f]$, the uplifted one is defined as the tensor in ${\cal T}[N_f+1]$ which is identical to $T$, but with indices running from 1 to $N_f+1$:
\bea
\label{eq:uplifting}
T^{\{n_L\}; \{n_R\}}_{\{\bar{n}_L\}; \{\bar{n}_R\}} \in {\cal T}[N_f] \to T^{\{n_L\}; \{n_R\}}_{\{\bar{n}_L\}; \{\bar{n}_R\}} \in {\cal T}[N_f+1]\, .
\eea
We notice that:
\begin{itemize}
\item For any tensor $T^{\{n_L\};\{n_R\}}_{\{\bar{n}_L\};\{\bar{n}_R\}} \in {\cal T}[N_f]$ there exists a corresponding well defined tensor $T^{\{n_L\};\{n_R\}}_{\{\bar{n}_L\};\{\bar{n}_R\}} \in {\cal T}[N_f+1]$. This is because the YTs $\{n_L\}$, $\{n_R\}$, $\{\bar{n}_L\}$ and $\{\bar{n}_R\}$ have at most $N_f$ rows, hence are well defined for a theory with $N_f+1$ flavors as well.
\item Uplifting is an injective map: Any two different tensors of $ {\cal T}[N_f]$ are mapped to different tensors of ${\cal T}[N_f+1]$.
\item Uplifting is not surjective, since there are tensors of ${\cal T}[N_f+1]$ which cannot be obtained from the uplifting of any tensor of ${\cal T}[N_f]$. In particular, uplifting cannot reach tensors that have $N_f+1$ fully antisymmetrized upper (or lower) indices forming a singlet, since such configuration does not exist for tensors of ${\cal T}[N_f]$. 
\end{itemize}
The cartoon of Fig.~\ref{fig:cartoon_uplifting} summarizes the properties of tensor uplifting.

Uplifting has been defined as a map between tensors, but it can be used to induce a corresponding map between  representations of ${\cal G}[N_f]$ and ${\cal G}[N_f+1]$. In order to do so, one has to fix the redundancy implied by the tensor notation and choose one representative tensor for each representation. A potential obstacle to a meaningful definition is however posed by the existence of equivalent tensors. It may happen, indeed, that two different tensors that are equivalent in a theory with $N_f$ flavors become inequivalent when the number of flavors is increased to $N_f+1$; in this case the same irrep of ${\cal G}[N_f]$ could be mapped to two different irreps of ${\cal G}[N_f+1]$, which is not acceptable. We know, on the other hand, that two class-A tensors cannot be equivalent, see Lemma~1 at the end of Section~\ref{sec:spectrum}. 
If we thus consider only representations of ${\cal G}[N_f]$ and ${\cal G}[N_f+1]$ that are interpolated by class-A tensors,  then a meaningful definition of uplifting is possible also for irreps. More explicitly: given any representation $r$ interpolated by a class-A tensor $T \in {\cal T}[N_f]$, the uplifted representation $U[r]$ is defined to be the one interpolated by the class-A tensor of ${\cal T}[N_f+1]$ which is identical to $T$. 
Since Lemma 1 ensures that two class-A tensors cannot be equivalent, there exists at most one class-A tensor interpolating any given representation, i.e. any given state.~\footnote{Class-A tensors interpolating a baryon will necessarily have $\bar n=0$, while exotics will be interpolated by class-A tensors with $\bar n \not = 0$.} This means that the above definition is unique and gives an injective mapping. 

\vspace{0.3cm}
\textbf{Example 4.2}. Let us consider a theory with $N_c=3$ and the following two tensors
\begin{equation}
T_a = T^{\{3\} ; \{ 0\}}_{\{0\} ; \{0\}} , \qquad T_b = T^{\{4\} ; \{ 0\}}_{\{1\} ; \{0\}}
\end{equation}
where $\{3\}$ and $\{4\}$ are YT with a single column with respectively 3 and 4 rows. For $N_f=4$, $T_a$ is class A and $T_b$ is class B; they transform under the same representation of ${\cal G}[4]$,
\begin{equation}
r_a = \left( \,\tiny\yng(1,1,1) \, ,s , 1 \right) = \left( \,\tiny\yng(2,2,2,1)\, ,s , 1 \right)  = r_b\, ,
\end{equation}
and are therefore equivalent. For $N_f=5$, $T_a$ and $T_b$ are still respectively of class A and B, but are now inequivalent since
\begin{equation}
r_a = \left( \,\tiny\yng(1,1,1)\, ,s , 1 \right) \not = \left( \,\tiny\yng(2,2,2,2)\, ,s , 1 \right)  = r_b\, ,
\end{equation}
Establishing an uplifting map between the representations of $N_f=4$ and $N_f=5$ theories is thus possible only by focusing on representations interpolated by class-A tensors, i.e. by considering $r_a$ and discarding $r_b$.
\vspace{0.3cm}

Notice that although every class-A irrep in ${\cal R}[N_f]$ can be uplifted to a class-A irrep in ${\cal R}[N_f+1]$, the converse is not true.~\footnote{Here and in the following, by class-A irrep we mean an irrep that is interpolated by a class-A tensor.} In particular, irreps in ${\cal R}[N_f+1]$ that are interpolated by tensors with $n+\bar n=N_f$ are not the uplift of any class-A irrep in ${\cal R}[N_f]$.

Taking one step further, the map between representations corresponds to a map between the physical massless states of a theory with $N_f$ flavors and those of a theory with $N_f+1$ flavors, provided that the latter theory has the confining description. Such a last step is however not necessary to establish $N_f$-independence, since PMC$[N_f]$ can be considered as mathematical equations rather than physical conditions.

Uplifting can be used to establish $N_f$-independence of $\text{PMC}[N_f,1]$. It is easy to see that $N_f$-independence holds if it is possible to establish an injective map between representations of ${\cal G}[N_f]$ and ${\cal G}[N_f+1]$, as well as between terms in their decompositions when one flavor is given a mass. The definition of uplifting given above has such a property, because Lemma~1 ensures that within class~A no equivalent tensors exist of ${\cal G}[N_f]$ and of ${\cal G}[N_f,1]$ as well.
More in detail, one can impose the following restriction on the spectrum of massless bound states (i.e. on the set of possible representations): only states (representations) that can be interpolated by a class-A tensor are assumed to have non-vanishing index.
Given a spectrum $\{ \ell(r)\}$ fulfilling this hypothesis, with $r\in {\cal R}[N_f]$, it is possible to define an uplifted spectrum
of representations $r'\in {\cal R}[N_f']$ as follows:
\begin{equation}
  \label{eq:Usol}
  \left\{ \ell(r^\prime)\right\} = U^{(N'_f-N_f)}\!\left[ \{ \ell(r)\}\right]: \qquad \ell(r^\prime) = \begin{cases} \ell(r) & \text{if } r' = U^{(N'_f-N_f)}[r] \\
  0 & \text{otherwise} \end{cases}\, ,
\end{equation}
where $U^{(N'_f-N_f)}[r]$ denotes the irrep of ${\cal R}[N'_f]$ obtained by uplifting $(N'_f-N_f)$ times the irrep $r$ of ${\cal R}[N_f]$.
One thus arrives at the following
\\[0.25cm] \noindent {\bf Lemma 2.} \ \textit{Let $\left\{ \ell\!\left(r\right)\right\}$ be a real solution of $\text{PMC}\,[N_f,1]$ such that the index of any state that cannot be interpolated by a class-A tensor vanishes. Then $U^{(N'_f-N_f)}\!\left[ \{ \ell(r)\}\right]$ is a real solution of $\text{PMC}\,[N'_f,1]$ for any $N_f' \geq N_f$.}
\\[0.4cm]
The proof follows easily from the definition of uplifting and the fact that it is an injective map.

\subsubsection{Condition for $N_f$-independence of AMC$\,[N_f]$}

We now turn to the $\text{AMC}[N_f]$. These equations have an explicit dependence on $N_f$ through the anomaly coefficients as well as an implicit one, since increasing the number of flavors gives rise to new tensor structures, hence new irreps. As for the PMC, one can consider Eq.~(\ref{eq:am_2}) as a rule to generate a set of $\text{AMC}[N_f]$ for any $N_f$, independently of the IR behavior of the corresponding theory. Then the following result holds true:
\\[0.35cm] \noindent {\bf Theorem 1 ($N_f$ independence).} \ \textit{Let $\left\{ \ell( r)\right\}$ be a real solution of $\text{AMC}\,[N_f]\cup\text{PMC}\,[N_f,1]$ such that the index of any state that cannot be interpolated by a class-A tensor vanishes. Then $U^{(N'_f-N_f)}\!\left[ \{ \ell(r)\}\right]$ is a real solution of $\text{AMC}\,[N'_f]\cup\text{PMC}\,[N'_f,1]$ for any $N'_f\geq N_f$.}
\\[0.5cm]
The proof is most easily obtained by induction. Let us start by considering the AMC for $N_f+1$ flavors,
\begin{equation}
\label{eq:AMCNf+1}
N_c =\sum_{r^\prime\in {\cal R}[N_f+1]} \,\ell\!\left(r^\prime\right) d\!\left(r'_R \right)  \bar D_{i}\!\left(r'_L\right)\, ,
\end{equation}
which we want to prove to hold for $\ell\!\left(r^\prime \right)$ defined in terms of $\ell\!\left(r\right)$ as specified by the definition of uplifted solution of Eq.~(\ref{eq:Usol}): $\{ \ell(r') \} = U\left[ \{ \ell(r) \} \right]$. In particular, the sum in Eq.~(\ref{eq:AMCNf+1}) runs over the representations $r' = (r'_L,r'_R,b)$ of ${\cal G}[N_f+1]$ that are the uplift of class-A representations $r=(r_L,r_R,b)$ of ${\cal G}[N_f]$. This means that any $r'$ itself is interpolated by a class-A tensor of ${\cal T}[N_f+1]$.  It is possible to decompose the dimension $d$ and the anomaly coefficient~$\bar D_i$ in terms of those of the CYT of $SU(N_f)$; one has:
\begin{align}
\label{eq:decd}
\begin{split}
d\!\left(r'_R\right) & = \sum_{r'_{1R}} k(r'_R \to r'_{1R}) \,  d(r'_{1R}) \, ,
\end{split} \\[0.4cm]
\label{eq:decA}
\begin{split}
\bar D_{i}\!\left(r'_L\right) & =\sum_{r'_{1L}}  k(r'_L \to r'_{1L}) \,  \bar D_{i}\!\left(r'_{1L}\right)\, ,
\end{split}
\end{align}
where the sums in Eqs.~(\ref{eq:decd}) and~(\ref{eq:decA}) run over all the irreps after decomposition.
By inserting Eqs.~(\ref{eq:decd}),~(\ref{eq:decA}) into (\ref{eq:AMCNf+1}) one obtains
\begin{equation}
\label{eq:intermediateAMC}
N_c =  \sum_{r'_{1R}} \sum_{r'_{1L}} d(r'_{1R}) \, \bar D_{i}\!\left(r'_{1L}\right) \sum_{r^\prime\in {\cal R}[N_f+1]} \ell\!\left(r^\prime \right) k\!\left( r^\prime \to r'_1 \right)\, ,
\end{equation}
where $r'_1 = \left(r'_{1L}, r'_{1R}, H_1, b \right)$ is a generic irrep of ${\cal G}[N_f+1,1]$.
By setting $r'_{1L} = \left\{m_L;\bar{m}_L\right\}$ and $r'_{1R} = \left\{m_R;\bar{m}_R\right\}$ one has $H_1=bN_c - (m_L+m_R) + (\bar{m}_L+\bar{m}_R)$, and 
\begin{equation}
  \label{eq:kappa}
  k\!\left( r^\prime \to r'_1 \right) = k(\left\{n_L; \bar{n}_L\right\} \to \left\{m_L;\bar{m}_L\right\}) \times
  k(\left\{n_R;\bar{n}_R\right\} \to \left\{m_R;\bar{m}_R\right\})\, .
\end{equation}
By Lemma~2, $\left\{ \ell\!\left(r^\prime\right)\right\}$ is a solution of PMC$[N_f+1,1]$, hence the last sum of Eq.~(\ref{eq:intermediateAMC}) vanishes unless $H_1=0$. One can thus rewrite Eq.~(\ref{eq:intermediateAMC}) as
\begin{equation}
\label{eq:finalAMC}
N_c = \sum_{r} d(r_R) \, \bar D_{i}\!\left(r_L\right) \sum_{r^\prime\in {\cal R}[N_f+1]} \ell\!\left(r^\prime \right) k\!\left( r^\prime \to r\right)\, ,
\end{equation}
where $r = \left(r_L, r_R, b \right)$ is now a representation of ${\cal R}[N_f]$ (since it belongs to the subset of representations $r'_1$ with $H_1=0$, and ${\cal R}_0[N_f+1,1]={\cal R}[N_f]$).

The meaning of Eq.~(\ref{eq:finalAMC}) can be better understood by introducing the following definition of downlifting: given a spectrum $\{ \ell(r')\}$ of irreps $r' \in {\cal R}[N_f+1]$, we define a downlifted spectrum of irreps $r\in {\cal R}[N_f]$ as
\begin{equation}
  \label{eq:Dsol}
  \left\{ \ell(r)\right\} = D\!\left[ \{ \ell(r')\}\right]: \qquad \ell(r) = \sum_{r^\prime\in {\cal R}[N_f+1]} \ell\!\left(r^\prime \right) k\!\left( r^\prime \to r\right)\, .
\end{equation}
By $k\!\left( r^\prime \to r\right)$ here we mean the multiplicity for the decomposition of $r'$ into the representation $r_1'$ with $H_1=0$ of ${\cal R}_0[N_f+1,1]$ which corresponds to $r$ by virtue of the identification ${\cal R}_0[N_f+1,1]={\cal R}[N_f]$. 
It is possible to show that, for any class-A $r \in {\cal R}[N_f]$,
\begin{equation}
  \label{eq:finalstep}
  \sum_{r^\prime\in {\cal R}[N_f+1]} \ell\!\left(r^\prime \right) k\!\left( r^\prime \to r\right) = \ell(U[r])\quad \qquad \text{that is:}
  \quad D[U[\{ \ell(r)\}]] = \{ \ell(r)\}\, .
\end{equation}
In other words: \textit{downlifting is the inverse of uplifting}.
Before providing the proof of this identity, we first discuss its implications.

The last sum in Eq.~(\ref{eq:finalAMC}) gives the downlift of the indices $\ell(r')$; since we defined the latter to be the uplift of the $\ell(r)$, from the result stated in (\ref{eq:finalstep}) it follows that Eq.~(\ref{eq:finalAMC}) coincides with AMC$[N_f]$, which is identically satisfied. This proves that $\{\ell\!\left(r^\prime \right)\}$, defined as specified in Theorem~1, is a solution of AMC$[N_f+1]$. Similarly, one can uplift the solution from $N_f+1$ to $N_f+2$ and so on. We thus obtain that $\text{AMC}[N_f]\cup\text{PMC}[N_f,1]$ implies $\text{AMC}[N'_f]\cup\text{PMC}[N'_f,1]$ for any $N'_f > N_f$.  This proves Theorem~1.

From Theorem 1, chiral symmetry breaking is proven as follows. Let us consider a theory with $N_f$ flavors in the confining regime and assume that integer solutions of $\text{AMC}[N_f]\cup\text{PMC}[N_f,1]$ exist which satisfy the following property: Only massless bound states interpolated by class-A tensors have non-vanishing indices.
Then Theorem 1 is valid and $N_f$ independence of $\text{AMC}[N_f]$ follows; by the argument presented at the beginning of this section, this in turn implies that no integer solution of $\text{AMC}[N_f]$ can exist, contradicting the initial assumption. This proves that either chiral symmetry is spontaneously broken, or the assumed property of the spectrum is violated.

The $N_f$-independence of solutions of $\text{AMC}[N_f]\cup\text{PMC}[N_f,1]$ is sufficient to prove chiral symmetry breaking. Under the same assumptions of Theorem~1, on the other hand,  it is possible to show that solutions of $\text{AMC}[N_f]\cup\text{PMC}[N_f]$ are also $N_f$-independent. One can thus state the following 
\\[0.35cm] \noindent {\bf Theorem 2 ($N_f$ independence - stronger version).} \ \textit{Let $\left\{ \ell( r)\right\}$ be a real solution of $\text{AMC}\,[N_f]\cup\text{PMC}\,[N_f]$ such that the index of any state that cannot be interpolated by a class-A tensor vanishes. Then $U^{(N'_f-N_f)}\!\left[ \{ \ell(r)\}\right]$ is a real solution of $\text{AMC}\,[N'_f]\cup\text{PMC}\,[N'_f]$ for any $N'_f\geq N_f$.}
\\[0.5cm]
For the proof, we need to show that $\{\ell\!\left(r^\prime \right)\} = U^{(N'_f-N_f)}\!\left[ \{ \ell(r)\}\right]$ is also a solution of PMC$[N'_f,i>1]$. Let us start by considering PMC$[N_f+1,2]$; these equations are of the form:
\begin{equation}
  0 = \ell( \hat r'_2) = \sum_{r'_1 \in {\cal R}_0[N_f+1,1]} k(r'_1 \to \hat r'_2) \ell(r'_1) \, , \qquad \forall \ \hat r'_2 \in \hat {\cal R}[N_f+1,2]\, ,
\end{equation}
where $\ell(r'_1)$ is the multiplicity of the irrep $r'_1$ with $H_1=0$ which appears in the decomposition of~$r'$:
\begin{equation}
  \label{eq:ellr1p}
\ell(r'_1) = \sum_{r' \in {\cal R}[N_f+1]} k(r' \to r'_1) \ell(r')\, .
\end{equation}  
Since each $r'_1$ in $ {\cal R}_0[N_f+1,1]$ can be identified with a corresponding irrep in $ {\cal R}[N_f]$, it is clear from Eq.~(\ref{eq:ellr1p}) that the $\ell(r_1')$ are just the downlift of the $\ell(r')$. From Eq.~(\ref{eq:finalstep}) and the definition of $\ell(r')$, it follows that $\{ \ell(r'_1)\} = D[\{\ell(r')\}] = D[U[\{\ell(r)\}]] = \{\ell(r)\}$.
We know from Section~\ref{sec:pmc} that PMC$[N_f+1,2]$ are identical to PMC$[N_f,1]$ after setting $\ell(r'_1) = \ell(r)$; therefore, this proves that $\{\ell(r') \}$ is a solution of PMC$[N_f+1,2]$. It is straightforward to show that $\{\ell(r') \}$ is also a solution of PMC$[N_f+1,i]$ for any $i>1$, since the latter equations coincide with PMC$[N_f,i-1]$. Similarly, by uplifting to higher values of $N_f$ one can easily show that $\{\ell\!\left(r^\prime \right)\}$ is a solution of PMC$[N'_f, i>1]$ for any $N'_f>N_f$. This proves Theorem~2.

Finally, we now provide the proof of Eq.~(\ref{eq:finalstep}).

\subsubsection{Proof of Eq.~(\ref{eq:finalstep}): downlifting is the inverse of uplifting}
\label{sec:singlets}

Let us start by noticing that if one assumes a purely baryonic spectrum, then Eq.~(\ref{eq:finalstep}) follows easily. Indeed, in that case $\bar n_{L,R}$ and $\bar m_{L,R}$ vanish in Eq.~(\ref{eq:kappa}), and $H_1=0$ implies $m_L = n_L$, $m_R=n_R$ (because $H_1=n_L+n_R - m_L-m_R$ and $m_L\leq n_L$, $m_R\leq n_R$). Therefore, $r'$ and $r$ are interpolated by the same tensor, that is: the only irrep~$r'$ contributing in the sum of Eq.~(\ref{eq:finalstep}) is $r' = U[r]$. Since $k(U[r]\to r)=1$, then Eq.~(\ref{eq:finalstep}) follows.
This justifies the results of the early works~\cite{Frishman:1980dq,Takeshita:1981sx,Kaul:1981fd}.

For a spectrum with exotics the proof is less straightforward and, to the best of our knowledge, cannot be found in previous literature.
Here we rely on the assumption that only bound states interpolated by class-A tensors have non-vanishing index. The spectrum of corresponding representations is finite, since the number of upper and lower indices of class-A tensors must satisfy $n+\bar n < N_f$. In particular, class-A tensors can have a maximum number of lower indices $\bar n_\text{max} < (N_f-bN_c)/2$. In this section we will assume $\bar n_\text{max} =1$ for simplicity,  the generalization to $\bar n_\text{max}>1$ is discussed in Appendix~\ref{app:exotics}.
Since baryon number is unchanged when decomposing a tensor, i.e. bound states of different baryon numbers do not mix in PMC, we work at fixed $b$ in the following.

For $\bar n_\text{max} =1$, there are only two kinds of traceless tensors of ${\cal T}[N_f+1]$ to be considered: 
\bea
\label{eq:form}
T^{\{b N_c+1\}}_{\{1\}} \quad \text{and} \quad T^{\{b N_c\}}_{\{0\}}\ .
\eea
Those interpolating the states $r'$ in Eq.~(\ref{eq:finalstep}) will have this form and, by the assumptions of Theorem~1, can be chosen to belong to class A. Tensors interpolating the irreps $r'_1$ of ${\cal G}[N_f+1,1]$ with $H_1 = i \geq 0$ will be of the form $T^{\{b N_c+1-i\}}_{\{1\}}$ and $T^{\{b N_c-i\}}_{\{0\}} $. In particular, those with $H_1=0$ are of the same form as in Eq.~(\ref{eq:form}).

First, we consider the PMC relative to irreps $\hat r'_1$ of ${\cal G}[N_f+1,1]$ that have $H_1=i >0$ and that are interpolated by tensors with $(bN_c+1-i)$ upper indices and 1 lower index. For each such representation there is one PMC of the form
\begin{equation}
\label{eq:typeA}
\begin{split}
0 & =  \ell\!\left( \hat r'_1\right) \\
& = \sum_{r^\prime \in {\cal R}[N_f+1]} \ell\!\left( r^\prime\right) k\!\left( T^{\{b N_c+1\}}_{\{1\}} \to T^{\{b N_c+1-i\}}_{\{1\}} \right) \\
& = \sum_{\{ bN_c+1 \} } \ell\!\left( r^\prime\right) k\!\left( \{bN_c+1\} \to \{bN_c+1-i\} \right)\, ,
\end{split}
\end{equation}
where $T^{\{b N_c+1\}}_{\{1\}}$ transforms as the representation $r'$ of ${\cal G}[N_f+1]$. Next, we compute the sum of Eq.~(\ref{eq:finalstep}) for representations $r$ (with $H_1=0$) that are interpolated by tensors $T^{\{b N_c\}}_{\{0\}}$. One has (using $k\!\left( \{1\} \to \{0\} \right)=1$):
\begin{equation}
  \begin{split}
\sum_{r^\prime} \ell\!\left(r^\prime \right) k\!\left( r^\prime \to r\right) =  
&\  \ell\!\left( U[r]\right) + \sum_{r'\not = U[r]} \ell\!\left( r'\right) k\!\left( T^{\{b N_c+1\}}_{\{1\}} \to T^{\{b N_c\}}_{\{0\}} \right) \\
= & \ \ell\!\left( U[r]\right)+  \sum_{ \{ bN_c+1 \} } \ell\!\left( r'\right) k\!\left( \{b N_c+1\} \to \{b N_c\} \right)  \, .
\end{split}
\end{equation}
The sum in the second line coincides with that appearing in the last line of Eq.~(\ref{eq:typeA}) for $i=1$; hence it vanishes and we obtain Eq.~(\ref{eq:finalstep}). Finally, Eq.~(\ref{eq:finalstep}) follows trivially if $r$ is interpolated by a tensor $T^{\{b N_c+1\}}_{\{1\}}$, since in this case $r'$ can only be interpolated by the same tensor, i.e. $r' = U[r]$. This concludes our proof for $\bar n_\text{max} =1$.

\section{Conclusions}\label{sec:conclusion}

In this work we have critically reconsidered the argument based on 't Hooft anomaly matching that aims at proving chiral symmetry breaking in confining QCD-like theories. The assumption that the theory is in the confining regime implies nontrivial constraints on the possible irreps under which hadrons transform under the chiral symmetry group.
The main line of reasoning, first suggested by 't Hooft~\cite{tHooft:1979rat} and later developed by many other authors~\cite{Frishman:1980dq, Schwimmer:1981yy, Farrar:1980sn, Takeshita:1981sx, Kaul:1981fd, Cohen:1981iz,Bars:1981nh}, relies on $N_f$-independence. Such property was assumed to hold in previous works based on qualitative arguments, and it was never proven rigorously.

In this paper we provided a proof of $N_f$-independence by establishing an injective map between massless bound states of theories with different number of flavors. We made use of tensor notation and found that such a map can be established only on  bound states that are interpolated from the vacuum by class-A tensors (see Sec.~\ref{sec:equivtensor} for a definition). We therefore conclude that $N_f$-independence holds, hence chiral symmetry breaking is proven, only by making a restriction on the possible massless bound states in the spectrum. Our analysis includes exotic bound states interpolated by tensor operators with lower (antiquark) indices, which were not considered in the past literature except for the work of Ref.~\cite{Farrar:1980sn}.

Another interesting strategy to prove chiral symmetry breaking using 't Hooft anomaly matching and $N_f$-independence was proposed by Cohen and Frishman in Ref.~\cite{Cohen:1981iz}. These authors focused on the purely baryonic case and suggested that, although AMC$[N_f]$ and PMC$[N_f,1]$ equations depend on $N_f$ in general, there might be one linear combination of them which does not. Such `$N_f$-equation' can be derived from AMC at large $N_f$ and, once evaluated at $N_f=0$ as justified by its $N_f$-independence, it has no integer solutions. Proving the existence of the $N_f$-equation is not straightforward, since the system of AMC$[N_f]$ and PMC$[N_f,1]$ does not remain the same as $N_f$ is varied, and the arguments provided in Ref.~\cite{Cohen:1981iz} are not conclusive. The best we can infer from our work is that the derivation of this equation is not in contradiction with the downlifting procedure used in Ref.~\cite{Ciambriello:2024xzd}.

The alternative approach proposed by Schwimmer in Ref.~\cite{Schwimmer:1981yy} makes use of the $SU(N_f|N_f)$ superalgebra and does not rely on $N_f$-independence.  As such, it is a promising candidate for a purely algebraic proof of chiral symmetry breaking.  We have explicitly verified that, in the case of QCD-like theories with $N_c=3$ and 5, the superalgebra solution coincides with the generic solution of PMC equations if one assumes a purely baryonic spectrum of massless states. This result holds for any number of flavors $N_f >2$, i.e. even in presence of equivalent tensors.
  It would be desirable to extend the work of Schwimmer to the case of a massless spectrum with exotics, and prove that a generic PMC solution can be always written in terms of irreps of $SU(N_f|N_f)$. In this way one would obtain an algebraic proof of chiral symmetry breaking valid for any $N_f >2$.  We leave such study to future work.

The notion of global symmetries has been recently generalized~\cite{Gaiotto:2014kfa}, providing some new perspective on strongly-coupled gauge dynamics. Among the many papers on this subject that appeared in the literature, those which use 't Hooft anomaly matching of generalized symmetries include studies analyzing the vacuum of gauge theories at non-vanishing theta angle~\cite{Gaiotto:2017yup,Gaiotto:2017tne}, constraining the dynamics in the space of coupling constants~\cite{Cordova:2019jnf,Cordova:2019uob}, constraining the phase diagram of QCD at finite temperature and/or finite density~\cite{Tanizaki:2017mtm}, excluding an exotic chiral symmetry breaking phase of QCD~\cite{Tanizaki:2018wtg}, constraining the infrared dynamics of QCD-like theories with quarks in higher-dimensional representations with non-trivial N-ality~\cite{Anber:2019nze} and of chiral gauge theories (see e.g. Ref.~\cite{Bolognesi:2021jzs} for a recent review and references therein). Such line of research is extremely interesting and might lead to additional insight on the occurrence of the ordinary chiral symmetry breaking phase in QCD-like theories where more traditional arguments like those used in this paper fail.

In this work we focused on non-supersymmetric vector-like gauge theories with fermions in the fundamental representation of the color group. It would be interesting to extend our results to theories with fermions in different representations, or with different gauge groups. Studies of chiral symmetry breaking in some of these theories have been made in the literature, but we expect that our approach can be used to obtain further progress in a more systematic way.

\section*{Acknowledgments}
We would like to thank Marco Bochicchio, Csaba Cs\'{a}ki, Ethan Neil, Mauro Papinutto, Slava Rychkov, Yael Shadmi and Giovanni Villadoro for useful discussions and comments.
This research was supported in part by the MIUR under contract 2017FMJFMW (PRIN2017) and performed in part at Aspen Center for Physics, which is supported by National Science Foundation grant PHY-2210452.
The work of R.C. was partly supported by a grant from the Simons Foundation and by the Munich Institute for Astro- and Particle Physics (MIAPP), which is funded by the Deutsche Forschungsgemeinschaft (DFG, German Research Foundation) under Germany Excellence Strategy - EXC-2094 – 390783311.
The work of A.L. was partially supported by the INFN special initiative grant “GAST” (Gauge and String Theories). 
L.X.X. would like to thank Scuola Normale Superiore in Pisa, where this project was initiated,  for its warm hospitality. L.X.X. was supported in part by the National Science Foundation of China under Grants No. 11635001 and No. 11875072 while he was working on this project at Peking University.

\appendix

\section{Technicalities on Anomalies}
\label{app:trace}
This Appendix further clarifies some technicalities encountered in the previous sections.

For quarks in the fundamental representation, the anomaly traces are
\begin{equation}
\begin{split}
D_{i=2}\!\left(\tiny\yng(1)\right) \delta_{a b} &= \frac{1}{N_c}\ \tr\!\!\left[\{\lambda^F_a, \lambda^F_b\}\right]\equiv \frac{1}{N_c} \delta_{a b}\, ,\\[0.2cm]
D_{i=3}\!\left(\tiny\yng(1)\right) d_{abc} &= \tr\!\!\left[\{\lambda^F_a, \lambda^F_b\} \lambda^F_c\right]\equiv d_{abc}  \, .
\end {split}
\end{equation}
Here $\lambda^F_{a}\ (a=1,2,\cdots, N_f^2-1)$ are the generators of $SU(N_f)$ in the fundamental representation, whose normalization satisfies $\tr\![\lambda^F_a\lambda^F_b]=\delta_{ab}/2$.
The anomaly traces for bound states that appear in Eq.~(\ref{eq:am}) are defined as 
\begin{equation}
\begin{split}  
D_{i=2}\!\left(r\right) \delta_{a b} &= b\ \tr\!\!\left[\{\lambda_a, \lambda_b\}\right]\, ,\\[0.2cm]
D_{i=3}\!\left(r\right) d_{abc} &= \tr\!\!\left[\{\lambda_a, \lambda_b\} \lambda_c\right] \, . 
\end{split}
\end{equation}
Here $\lambda_{a}\ (a=1,2,\cdots, N_f^2-1)$ are the generators of $SU(N_f)$ in the representation $r$.
The constants $\bar D_{i}\!\left( r\right)$ are conveniently defined as the ratios of anomaly traces, see Eq.~(\ref{eq:ano_const}). We notice that the values of $\bar D_{i}\!\left( r\right)$ do not depend on which generators are chosen in calculating the traces; without loss of generality, one can choose generators belonging to the $SU(N_f-1)$ subalgebra.

It is worthwhile to comment more on Eq.~(\ref{eq:decd}) and Eq.~(\ref{eq:decA}), given their importance in the proof of $N_f$-independence. Equation~(\ref{eq:decd}) just reflects the fact that the dimensionality of a representation is equal to the sum of the dimensionalities of the representations appearing in its decomposition. Equation~(\ref{eq:decA}) is less trivial to justify, and we do so by considering the following facts:
\begin{enumerate}
\item The constants $\bar D_{i}\!\left( r'\right)$, where $r'$ is an irrep of $SU(N_f+1)$, can be calculated by using the generators belonging to the $SU(N_f)$ subalgebra of the original $SU(N_f+1)$; these act nontrivially on $r'$ only when its indices are charged under $SU(N_f)$. This can be schematically indicated by
\bea
\bar D_{i}\!\left( r' \right)=\bar D_{i}\!\left( r'\right)|_{SU(N_f)}\, ,
\label{eq:app_a_1}
\eea
where on the right-hand side we mean that $\bar D_{i}$ can be evaluated by using the $SU(N_f)$ subalgebra.
\item The irrep $r'$ is reducible under the $SU(N_f)$ subalgebra, i.e. it can be decomposed into a series of irreducible representations $(r, H)$ of $SU(N_f)\times U(1)_H$,
\bea
r'=\bigoplus\limits_{r} \left(r, H\right)\, ,
\eea
where each index of $r'$ varies from $1$ to $N_f+1$, while each index of $r$ goes from $1$ to $N_f$. The $U(1)_H$ charge of the decomposed representations is given by $H=(n-m)-(\bar{n}-\bar{m})$, where $n$ and $\bar n$ ($m$ and $\bar m$) are the numbers of upper and lower indices of $r'$ (of $r$). The YTs characterizing $r'$ and $r$ can be identical, in which case $H=0$. It follows that
\bea
\bar D_{i}\!\left( r'\right)|_{SU(N_f)}=\bar D_{i}\!\left(\bigoplus\limits_{r} r \right)\, .
\label{eq:app_a_2}
\eea
\item Equation~(\ref{eq:decA}) follows from Eqs.~(\ref{eq:app_a_1}),(\ref{eq:app_a_2}) and the identity 
\bea
\bar D_{i}\!\left( r_1\bigoplus r_2\right)= \bar D_{i}\!\left( r_1\right)+ \bar D_{i}\!\left( r_2\right)
\eea
valid for any two irreps $r_1$ and $r_2$ of $SU(N_f)$. 
\end{enumerate}

\section{Equivalent Tensors}
\label{app:eq_tensor}
This Appendix contains the proof of Lemma~1 on the absence of equivalent tensors stated in Sec.~\ref{sec:spectrum}. Lemma~1 applies to class-A tensors, for which
\begin{equation}
\label{eq:equivtensors}
n+\bar n < N_f \, .
\end{equation}  
In particular, we will prove that this is a necessary and sufficient condition for the absence of equivalent tensors, hence defining the largest set of non-equivalent tensors.
\par As explained in Sec.~\ref{sec:spectrum}, one way to construct equivalent traceless tensors of ${\cal T}[N_f]$ is to let their YTs differ by the presence of one or more sets of $N_f$ fully antisymmetrized indices (flavor singlets). In this case, the CYTs also differ by columns with $N_f$ boxes. Clearly, for $n+\bar n < N_f$ no such tensors can arise. This condition also ensures that no equivalent traceless tensors arise after decomposition which transform under the same representation of ${\cal G}[N_f,1]$ and whose YT differ only by flavor singlets.

We thus consider the less trivial case where the equivalence arises by combining the YT of upper and lower indices. For simplicity, we will focus on tensors with only left or right indices, the generalization to tensors with both kinds of indices is straightforward. Let
\begin{equation}
\left\{ T^{\{n_1\}}_{\{\bar{n}_1\}}, T^{\{n_2\}}_{\{\bar{n}_2\}}, \cdots, T^{\{n_\kappa\}}_{\{\bar{n}_\kappa\}}\right\}
\end{equation}
be a class of $\kappa$ equivalent traceless tensors of this kind. Since no flavor singlets can arise due to Eq.~(\ref{eq:equivtensors}), all these tensors must have the same CYT. Let $\mathcal{R}=\mathcal{P}+\mathcal{Q}$ be the number of columns of the CYT, where $\mathcal{P}$ of them originate from the YT of upper (quark) indices, and the remaining $\mathcal{Q}$  from the dual YT of lower (antiquark) indices. For a given tensor $T^{\{n_j\}}_{\{\bar{n}_j\}}$, we will indicate with $\zeta^{(j)}_i$ the length of the $i$-th column of $\{n_j\}$, for $i= 1,\dots, \mathcal{P}$, and with $\tilde \zeta^{(j)}_i$ the length of the $i$-th column of $\widetilde{\{\bar{n}_j\}}$, for $i= 1,\dots, \mathcal{Q}$. The total number of boxes in the CYT then equals 
\begin{equation}
\label{eq:totboxes}
n_\text{box} = \sum_{i=1}^{\cal P} \zeta^{(j)}_i + \sum_{i=1}^{\cal Q} \tilde\zeta^{(j)}_i = n_j+\mathcal{Q} N_f -\bar{n}_j =b N_c+\mathcal{Q} N_f\, , 
\end{equation}
where the last equality follows from the sum rule~(\ref{eq:cons}). Since $n_\text{box}$ does not depend on $j$, all tensors in the equivalent class have the same $\mathcal{Q}$, and thus the same $\mathcal{P}$ as well. In other words, the class is characterized by $({\cal P},{\cal Q})$ and by the baryon number $b$. The number of equivalent tensors in a class $({\cal P},{\cal Q})_b$ is given by the binomial coefficient: 
\begin{equation}
\kappa = C_\mathcal{R}^\mathcal{P}=C_\mathcal{R}^\mathcal{Q}\equiv \frac{\mathcal{R} !}{\mathcal{P} ! \mathcal{Q} !}\, .
\end{equation}
Indeed, starting with a total of $\cal R$ columns in the CYT, each tensor corresponds to a different way of assigning $\cal P$ of them to the YT of upper indices (so that the remaining ones are columns of the dual YT of lower indices); there are $C_\mathcal{R}^\mathcal{P}$ such possible assignments.

By defining $n_{max} = \max_j \{ n_j\}$ and $\bar n_{max} = \max_j \{ \bar n_j\}$ as respectively the maximum number of quarks and antiquarks of any tensor in the class, these two inequalities follow:
\begin{align}
\label{eq:ineq1}
\sum^\kappa_{j=1} \bar{n}_j & \leq \kappa \, \bar{n}_{max}=C_\mathcal{R}^\mathcal{P} \, \bar{n}_{max}\\
\label{eq:ineq2}
\sum^\kappa_{j=1} n_j &\leq \kappa \, n_{max}=C_\mathcal{R}^\mathcal{P} \, (\bar{n}_{max}+b N_c)\ .
\end{align}
Furthermore, the following identity holds:
\begin{equation}
\label{eq:appen_1}
\mathcal{P} \sum^\kappa_{j=1} \bar{n}_j+\mathcal{Q} \sum^\kappa_{j=1} n_j= C_\mathcal{R}^\mathcal{P}\, \mathcal{P}\, \mathcal{Q} \, N_f\ .
\end{equation}
We postpone the proof to the end and focus first on its implications. Using (\ref{eq:appen_1}) together with (\ref{eq:ineq1}) and (\ref{eq:ineq2}), one obtains
\begin{equation}
\label{eq:ineq3}
N_f\leq \frac{\bar{n}_{max}}{\mathcal{Q}}+\frac{\bar{n}_{max}+b N_c}{\mathcal{P}}\, .
\end{equation}
For fixed $N_f$, the weakest lower bound on $\bar n_{max}$ and $b$ comes from the classes with $\mathcal{P}=\mathcal{Q}=1$ and reads $N_f \leq 2\bar{n}_{max}+b N_c$. Any class of this kind with $\mathcal{P}=\mathcal{Q}=1$ contains a pair of equivalent tensors ($\kappa=2$) with numbers of quarks and antiquarks related as follows: $n_1 = N_f - \bar n_2$, $n_2 = N_f - \bar n_1$.~\footnote{These identities follow by requiring that the CYT be the same for the two tensors. Indeed, the CYT has two columns. In the case of the first tensor, the first column originates from its quark indices and the second column originates from its antiquark indices, while for the second tensor the role of quark and antiquark indices is exchanged.}. The Eq. (\ref{eq:ineq3}) implies that either $2\bar n_1+bN_c\leq N_f$ or
$2\bar n_2+bN_c\leq N_f$. Therefore, we see that Eq.~(\ref{eq:equivtensors}) is a necessary requirement for avoiding pairs of equivalent tensors.
We notice that in fact no equivalent traceless tensors with identical CYTs can arise even for $2\bar n +bN_c = N_f$, when $\bar{n}\neq
0$. Indeed, if the inequality of Eq.~(\ref{eq:ineq3}) is saturated, then the second tensor has (as a consequence of the relations written above) the same $n$ and $\bar n$. In other words, in this case the two equivalent tensors are actually identical. The stronger condition of  Eq.~(\ref{eq:equivtensors}) is on the other hand required to exclude the existence of equivalent tensors whose YTs differ by flavor singlets.
\par The final step is to prove that Eq.~(\ref{eq:equivtensors}) is a sufficient condition for the absence of equivalent tensors. Let us consider a generic pair of equivalent tensors $T^{\{n_1\}}_{\{\bar{n}_1\}}, T^{\{n_2\}}_{\{\bar{n}_2\}}$ and suppose $n_1+\bar n_1<N_f$. Since they have the same $\mathcal{P}$ and $\mathcal{Q}$ they will be connected by exchanging the same number $r$ of columns of quarks and antiquarks. If we call $\delta$ and $\bar\delta$ respectively the number of quarks and antiquarks of $T^{\{n_1\}}_{\{\bar{n}_1\}}$ that were exchanged, we have that $n_2+\bar n_2=n_1+(rN_f-\bar\delta)-\delta+\bar n_1+(rN_f-\delta)-\bar\delta$. Since $\delta+\bar\delta\leq n_1+\bar n_1<N_f$, this imply that $n_2+\bar n_2>(2r-1)N_f\geq N_f$. This means that there cannot be equivalent tensors both satisfying Eq.~(\ref{eq:equivtensors}) and concludes our proof. 
\par A few additional considerations show that when Eq.~(\ref{eq:equivtensors}) is satified, no equivalent traceless tensors of ${\cal G}[N_f,1]$ can possibly arise as well. The argument leading to Eq.~(\ref{eq:ineq3}) can be repeated for traceless tensors of ${\cal G}[N_f,1]$ by replacing $n_j$ and $\bar n_j$ with the number of massless quarks $n_j'$ and $\bar n_j'$ characterizing a tensor of the class. The values of $n_j'$ and $\bar n_j'$ satisfy the sum rule
\begin{equation}
n_j' - \bar n_j' = b N_c - H\, , 
\end{equation}
where $H$ is the number of massive quarks minus the number of massive antiquarks in the bound state. Since $H$ corresponds to the charge under the $U(1)_H$ factor inside ${\cal G}[N_f,1]$, all tensors in the same class must have the same $H$. Since for every tensor $n_j' + \bar n'_j \leq n_j +\bar n_j -1$, then Eq.~(\ref{eq:equivtensors}) ensures that no singlets of $SU(N_f-1)$ can appear in either of the upper and lower YT of the decomposed tensor. The steps that led to Eq.~(\ref{eq:ineq3}) now imply
\begin{equation}
\label{eq:ineq4}
N_f -1 \leq \frac{\bar{n}'_{max}}{\mathcal{Q}}+\frac{n'_{max}}{\mathcal{P}}\, .
\end{equation}
The strongest constraint is obtained by setting $\mathcal{P}=\mathcal{Q}=1$. 
\begin{itemize}
	\item If $H>0$, then there are more massive quarks than antiquarks, and $n'_{max} \leq n_{max}-1$, $\bar n'_{max} \leq \bar n_{max}$. 
	\item Similarly, if $H<0$ there are more massive antiquarks than quarks, and $n'_{max} \leq n_{max}$,  $\bar n'_{max} \leq \bar n_{max}-1$. 
	\item Finally, if $H=0$ there is the same number of massive quarks and antiquarks; hence $n'_{max} \leq n_{max}-1$ and $\bar n'_{max} \leq \bar n_{max}-1$. 
\end{itemize}
From these inequalities and Eq.~(\ref{eq:ineq4}) it follows $N_f - 1\leq 2 \bar n_{max}+bN_c -1$, so that no equivalent traceless tensors of ${\cal G}[N_f,1]$ exist if Eq.~(\ref{eq:equivtensors}) is satisfied. Considerations similar to those made after Eq.~(\ref{eq:ineq3}) show that no equivalent tensors can arise with identical CYT even when $N_f = 2 \bar n_{max}+bN_c$; this limit case must be however excluded to avoid the presence of equivalent tensors whose YT differ by flavor singlets.

To complete our proof we have to establish the validity of Eq.~(\ref{eq:appen_1}). We start by considering the following sums:
\begin{align}
\label{eq:sum1}
\sum_{j=1}^\kappa n_j & = \sum_{j=1}^\kappa \sum_{i = 1}^{\cal P} \zeta^{(j)}_i \\
\label{eq:sum2}
\sum_{j=1}^\kappa \bar n_j & = \sum_{j=1}^\kappa \sum_{i = 1}^{\cal Q} (N_f - \tilde \zeta^{(j)}_i) = C_{\cal R}^{\cal P} {\cal Q} N_f - \sum_{j=1}^\kappa \sum_{i = 1}^{\cal Q}\tilde \zeta^{(j)}_i\, .
\end{align}
It is not difficult to perform the sums on the right-hand sides of Eqs.(\ref{eq:sum1}), (\ref{eq:sum2}). For example, $\sum_{j=1}^\kappa \sum_{i = 1}^{\cal P} \zeta^{(j)}_i$ can be computed by counting how many times a given column of length $\zeta$ appears in all possible YTs of upper indices within the class. This number is equal to the binomial coefficient $C^{{\cal P}-1}_{{\cal R}-1}$, which corresponds to the number of YTs one can form with that column and other ${\cal P}-1$ chosen from the ${\cal R} -1$ available in the CYT. This multiplicity does not depend on which column is considered initially. Similarly, the sum $\sum_{j=1}^\kappa \sum_{i = 1}^{\cal Q}\tilde \zeta^{(j)}_i$ is computed by noticing that a given column of length $\tilde \zeta$ appears $C^{{\cal Q}-1}_{{\cal R}-1}$ times in all possible dual YTs of lower indices; again, this multiplicity does not depend on the column considered. One thus obtains
\begin{align}
\label{eq:sum3}
\sum_{j=1}^\kappa \sum_{i = 1}^{\cal P} \zeta^{(j)}_i &= C^{{\cal P}-1}_{{\cal R}-1}\, n_\text{box} \\
\label{eq:sum4}
\sum_{j=1}^\kappa \sum_{i = 1}^{\cal Q} \tilde\zeta^{(j)}_i &= C^{{\cal Q}-1}_{{\cal R}-1}\, n_\text{box} \, ,
\end{align}
where $n_\text{box}$ is the total number of boxes in the CYT, see Eq.~(\ref{eq:totboxes}). Since
\begin{equation}
{\cal P} \, C^{{\cal Q}-1}_{{\cal R}-1} = \frac{({\cal R}-1)!}{({\cal P}-1)!({\cal Q}-1)!} = {\cal Q} \, C^{{\cal P}-1}_{{\cal R}-1}\, ,
\end{equation}
from Eqs.~(\ref{eq:sum1})-(\ref{eq:sum4}) it follows Eq.~(\ref{eq:appen_1}).

	\section{PMC with equal masses}
	\label{app:PMCsamemasses}
	
        In this Appendix, we argue that the PMC that one obtains by considering $k$ massive quarks with two or more equal masses are implied by PMC$[N_f, 1]$, $\dots$, PMC$[N_f, k]$. First, we prove the statement for two equal masses, and then we sketch how to extend the proof to the general case. Our proof relies on the fact that the spectrum of massless fermions with non-zero index is finite, i.e. there may be an arbitrary though finite number of irreps $r\in {\cal R}[N_f]$ such that $\ell(r) \not = 0$.~\footnote{In a 4D quantum field theory, the $a$-theorem~\cite{Komargodski:2011vj} implies an upper bound on the number of massless degrees of freedom in the IR in terms of the number of UV degrees of freedom, see also~\cite{Appelquist:1999hr}.}

If one gives two flavors the same mass, $\mathcal{G}[N_f, 2]$ is enlarged to~\footnote{Here we are ignoring, as done before in this paper, any discrete identification. In particular, the actual symmetry acting on the massive flavors is $U(2)_H = (SU(2)_H\times U(1)_H)/\mathbb{Z}_2$.}
\begin{equation}
\mathcal{G}[N_f, \underline{2}]=SU(N_f-2)_L \times SU(N_f-2)_R\times U(1)_{B} \times SU(2)_{H}\times U(1)_H\;\ ,
\end{equation}
where $SU(2)_H$ acts on the doublet of massive quarks, and the $U(1)_H$ symmetry counts their overall number. Here and in the following we underline quantities related to the two equal masses. Clearly, $\mathcal{G}[N_f, 2]$ is a subgroup of $\mathcal{G}[N_f, \underline{2}]$.
	
It is useful to denote an irrep of $\mathcal{G}[N_f, \underline{2}]$ with a tuple, $\{A, h, \underline{j}\}$,  where $A$ is an $SU(N_f-2)_L\times SU(N_f-2)_R\times U(1)_{B}$ representation, $h$ is the $U(1)_H$ charge, and $\underline{j}$ is an $SU(2)_H$ irrep with spin~$j$.
If $h$ is even (odd), then $j$ must be integer (half-integer); to see this, just notice that an operator interpolating $\{A, h, \underline{j}\}$ has a number of massive quarks equal to $h$, up to an arbitrary number of massive quark-antiquark pairs.~\footnote{From a more abstract viewpoint, the overlap between the $SU(2)_H$ center and $U(1)_H$ ensures that $h+2j=0 \mod 2$.}
We further denote an irrep of $\mathcal{G}[N_f, 2]$ with a tuple $\{A, h, m_j\}$. Here $h$ and $m_j$ codify the charges under $U(1)_{H_1}$ and $U(1)_{H_2}$ as $h_1=\frac{1}{2}h+m_j$ and $h_2=\frac{1}{2}h-m_j$, respectively. Notice that $h_1$ and $h_2$ are always integers due to the connection between $h$ and $j$.
By using the Vafa-Witten theorem (the same argument used in Section.~\ref{sec:pmc}), one can prove that
\begin{equation}
\sum_{r\in {\cal R}[N_f]} \ell(r)\; k(r \to \{A, h, \underline{j}\})=0\; 
\end{equation}
when $h$ and $j$ do not vanish simultaneously.
Our goal is to show that these equations can be obtained from PMC$[N_f, 1]$ and PMC$[N_f, 2]$. 
	
The first step is to notice that PMC$[N_f, 1]$ and PMC$[N_f, 2]$ imply that   
\begin{equation}
\sum_{r\in {\cal R}[N_f]} \ell(r)\; k(r \to \{A, h, m_j\})=0\;. \label{eq:pmc2a}
\end{equation}
for any irrep $\{A, h, m_j\}$ of $\mathcal{G}[N_f, 2]$  with $m_j$ and $h$ not simultaneously equal to zero. Let us call this set of equations $\overline{\text{PMC}}[N_f, 2]$. We assume that the spectrum of irreps $r$ with non-vanishing index is finite, i.e. the sum in Eq.~(\ref{eq:pmc2a}) is over a finite number of terms. Then it follows that the set of irreps $\{A, h, \underline{j}\}$ of ${\cal G}[N_f, \underline{2}]$ obtained from the decomposition of the original massless spectrum is also finite.
Let us fix $A$ and $h$ and call $\underline{j}_{max}$ the highest spin of $SU(2)_H$ that appears in the decomposition.
In general, we can decompose each $\{A, h, \underline{j}\}$ further in irreps of $\mathcal{G}[N_f, 2]$ as 
\begin{equation}
 \label{eq:decomposition}
\{A, h, \underline{j}\} = \{A, h, j\}\oplus \{A, h, j-1\} \oplus ... \oplus \{A, h, -j\}\, . 
\end{equation}

Now we can proceed with a `peeling off' argument. The decomposition (\ref{eq:decomposition}) implies that, being $j_{max}$ the maximal $SU(2)_H$ spin in our spectrum, $\{A, h, j_{max}\}$ appears only in the decomposition of $\{A, h, \underline{j}_{max}\}$, so
\begin{equation}
\sum_{r\in {\cal R}(N_f)} \ell(r)\, k(r\to \{A, h, \underline{j}_{max}\}) 
=  \sum_{r\in {\cal R}(N_f)} \ell(r)\,  k(r\to \{A, h, j_{max}\})=0\;. \label{eq:firstsheet}
\end{equation}
This is the PMC$[N_f, \underline{2}]$ equation for $\{A, h, \underline{j}_{max}\}$,  and we are done with the first layer.
	
Let us consider the second layer. The irrep $\{A, h, j_{max}-1\}$ appears in the decomposition of $\{A, h, \underline{j}_{max}\}$ and $\{A, h, \underline{j_{max}-1}\}$, in both cases only once. Hence,
\begin{equation}
\label{eq:line1}
  \begin{split}
&\sum_{r\in \mathcal{R}(N_f)} \ell(r) \; k(r\to \{A, h, \underline{j_{max}}\})  + \sum_{r\in \mathcal{R}(N_f)} \ell(r) \; k(r\to \{A, h, \underline{j_{max}-1}\}) \\
& = \sum_{r\in \mathcal{R}(N_f)} \ell(r) \; k(r\to \{A, h, j_{max}-1\}) =0\;.
\end{split}
\end{equation}
The first term in the first line of (\ref{eq:line1}) vanishes because of Eq.~(\ref{eq:firstsheet}), leaving the PMC$[N_f, \underline{2}]$ equation for $\{A, h, \underline{j_{max}-1}\}$. 
	
One can apply this same argument iteratively, obtaining the PMC equation for $\{A, h, \underline{j_{max}-i}\}$, with $i<j_{max}$, from the $\overline{\text{PMC}}[N_f, 2]$ equation for $\{A, h, j_{max}-i\}$ and the PMC$[N_f, \underline{2}]$ equations for $\{A, h, \underline{j_{max}-k}\}$ with $k<i$. If $h$ is even and non-vanishing, then one can proceed further and prove also the PMC equation for $\{A, h, \underline{0}\}$ by using the PMC equation for $\{A, h, 0\}$.
We have thus proven that the PMC$[N_f, \underline{2}]$ can be derived from PMC$[N_f, 1]$ and PMC$[N_f, 2]$ (more precisely, from $\overline{\text{PMC}}[N_f, 2]$). Clearly, it is also true that $\overline{\text{PMC}}[N_f, 2]$ can be derived from PMC$[N_f, \underline{2}]$. Indeed, 
\begin{equation}
\sum_{r\in {\cal R}(N_f)} \ell(r) k(r \to \{A, h, j\})=\sum_{j'\geq j}\, \sum_{r\in {\cal R}(N_f)} \ell(r) k(r \to \{A, h, \underline{j'}\})=0\;.
\end{equation}

The first straightforward generalization of this proof is to consider $k$ equal masses. In this case, the symmetry group is~\footnote{Again, we ignore the discrete identification in the sketch of the proof.}
\begin{equation}
  G[N_f, \underline{k}]=SU(N_f-k)\times SU(N_f-k) \times U(1)_{B} \times SU(k)_H \times U(1)_H
\end{equation}
$G[N_f, k]$ is a subgroup of $G[N_f, \underline{k}]$. In particular one can identify $U(1)_{H_1} \times U(1)_{H_2} \times ...\times U(1)_{H_k}$ with $U(1)_H \times U(1)^{k-1}$, where $U(1)^{k-1}$ is the Cartan subgroup of $SU(k)_H$.  
We can generalize the notation $\{A, h, \underline{j}\}$ used before with $\{A, h, \underline{\vec{\mu}}\}$, where we label the $SU(k)_H$ irrep with its highest weight $\vec{\mu}=(\mu_1, ..., \mu_{k-1})$. We remind that: \textit{i)} the $\mu_i$ are eigenvalues of the Cartan operators; \textit{ii)} it is possible to define an ordering between weight vectors, e.g. the lexicographic one: $\vec{\mu}>\vec{\mu}'$ if the first non-zero element of $\vec{\mu}-\vec{\mu}'$ is positive. As in the $k=2$ case, we further label the irreps $\{A, h, \vec{\mu}\}$ of $\mathcal{G}[N_f, k]$ with their $U(1)_H \times U(1)^{k-1}$ charges, $h$ and $\vec{\mu}$, instead of the $U(1)_{H_1}\times...\times U(1)_{H_k}$ ones. 
Upon decomposing an irrep of $\mathcal{G}[N_f, \underline{k}]$ into irreps of $\mathcal{G}[N_f, k]$, one obtains
\begin{equation}
\{A, h, \underline{\vec{\mu}}\} = \{A, h, \vec{\mu}\} \; \bigoplus_i \;  \{A, h, \vec{\mu}-\vec{\alpha}_i\} \;\oplus\; \dots
\end{equation}
Here $\vec{\alpha}_i$ are the positive roots of the $su(k)$ algebra, and the dots denote terms obtained by further subtractions of the $\vec{\alpha}_i$.
	
We can repeat the `peeling off' argument given before. Let us fix $A$ and $h$ and single out, in the (finite) space of all the irreps of $\mathcal{G}[N_f, \underline{k}]$, the one with the highest vector $\underline{\vec \mu}_{max}$.
Since the unique irrep $\{A, h, \vec{\mu}_{max}\}$ in the decomposed spectrum comes from the decomposition of $\{A, h, \underline{\vec{\mu}}_{max}\}$,~\footnote{The maximal weight vector, i.e. the element of the irrep space with the maximal weight, is unique in any $SU(k)$ irrep (see \cite{Georgi:1999wka}).} one has
\begin{equation}
  \label{eq:PMCmumax}
\sum_{r \in {\cal R}[N_f]} \ell(r) k(r \to \{A, h, \underline{\vec{\mu}}_{max}\}) =\sum_{r \in {\cal R}[N_f]} \ell(r) k(r \to \{A, h, \vec{\mu}_{max}\})=0\;. 
\end{equation}
This is the PMC$[N_f, \underline{k}]$ for $\{A, h, \underline{\vec{\mu}}_{max}\}$.

We can thus proceed to the second layer, i.e. the second highest weight in the spectrum, $\vec{\mu}_{max}'$. In general, this state can appear only in the decomposition of $\underline{\vec{\mu}_{max}}$ and of $\underline{\vec{\mu}_{max}'}$. Therefore, we have
\begin{equation}
  \begin{split}
0=& \sum_{r \in {\cal R}[N_f]} \ell(r)\, k(r \to \{A, h, \vec{\mu}_{max}'\}) \\
= & \sum_{r \in {\cal R}[N_f]} \ell(r)\, k(r \to \{A, h, \underline{\vec{\mu}}_{max}\})\, k(\{A, h, \underline{\vec{\mu}}_{max}\} \to \{A, h, \vec{\mu}_{max}'\}) \\
 & +\sum_{r \in {\cal R}[N_f]} \ell(r) \, k(r \to \{A, h, \underline{\vec{\mu}}_{max}'\})\, ,
\end{split}
\end{equation} 
where we used that $k(\{A, h, \underline{\vec{\mu}}_{max}'\} \to \{A, h, \vec{\mu}_{max}'\})=1$. The sum in the second line vanishes because of Eq.~(\ref{eq:PMCmumax}) and we thus obtain the PMC$[N_f, \underline{k}]$ for $\{A, h, \underline{\vec{\mu}}_{max}'\}$. 
	
Similarly to the $k=2$ case, one can proceed in this way layer by layer until $\vec{\mu}=0$. Again, the case $\vec{\mu}=0$ can be included only if $h\neq 0$.

In the most generic case, one has several blocks of flavors with equal masses. It is possible to generalize the proof in this case by proceeding block by block. One can start from the first block of degenerate quarks and prove that one obtains the same PMC equations if this degeneracy is broken. Then, one can proceed inductively with the remaining blocks.

\section{General Proof of Eq.~(\ref{eq:finalstep})}
\label{app:exotics}
In this Appendix we generalize the proof of Eq.~(\ref{eq:finalstep}) given in Section~\ref{sec:singlets} by considering a generic spectrum of massless bound states. The assumption is that each irrep with nonvanishing index can be interpolated by a class-A tensor, i.e. by a traceless tensor $T^{\{n\}}_{\{\bar n\}}$ with $n+\bar n = 2\bar n +bN_c < N_f$. This implies a finite maximal number of antiquarks $\bar n_{\text{max}} < (N_f- bN_c)/2$.
The strategy is to decompose tensors starting from those with the highest rank.

We first consider representations $r\in {\cal R}[N_f]$ interpolated by tensors with the highest rank, i.e. tensors  $T^{\{b N_c+\bar n_{\text{max}}\}}_{\{\bar n_{\text{max}}\}}$. In this case, only the representation $r'$ interpolated by the same tensor can appear in the sum of Eq.~(\ref{eq:finalstep}), i.e. $r' = U[r]$. Hence, Eq.~(\ref{eq:finalstep}) follows easily.

Next, we consider the PMC relative to irreps $\hat r'_1$ of ${\cal G}[N_f+1,1]$ interpolated by tensors with $(bN_c+\bar n_{\text{max}}-i)$ upper indices and $\bar n_{\text{max}}$ lower indices, i.e. $T^{\{b N_c+\bar n_{\text{max}}-i\}}_{\{\bar n_{\text{max}}\}}$. Here $i>0$ and the $U(1)_{H_1}$ charge of the irrep is $H_{1}=i>0$. For each such irrep of ${\cal G}[N_f+1,1]$ one has a PMC of the form
\begin{equation}
\label{eq:typeA_app}
\begin{split}
0 & =  \ell\!\left( \hat r'_1 \right) \\
& = \sum_{r'\in {\cal R}[N_f+1]} \ell\!\left( r'\right) k\!\left( T^{\{b N_c+\bar n_{\text{max}}\}}_{\{\bar n_{\text{max}}\}} \to T^{\{b N_c+\bar n_{\text{max}}-i\}}_{\{\bar n_{\text{max}}\}} \right) \\
& = \sum_{\{ bN_c+\bar n_{\text{max}} \} } \ell\!\left( r' \right) k\!\left( \{bN_c+\bar n_{\text{max}}\} \to \{bN_c+\bar n_{\text{max}}-i\} \right)\, .
\end{split}
\end{equation}
We thus compute the sum in Eq.~(\ref{eq:finalstep}) for representations $r$ (with $H_1=0$) interpolated by tensors $T^{\{b N_c+\bar n_{\text{max}}-1\}}_{\{\bar n_{\text{max}}-1\}}$. By making use of Eq.~(\ref{eq:typeA_app}) with $i=1$ we find
\begin{equation}
\label{eq:typeB_app}
\begin{split}
\sum_{r^\prime} \ell\!\left(r^\prime \right) k\!\left( r^\prime \to r\right)   
=& \  \ell\!\left( U[r]\right) + \sum_{r' \not = U[r] } \ell\!\left( r'\right) k\!\left( T^{\{b N_c+\bar n_{\text{max}}\}}_{\{\bar n_{\text{max}}\}} \to T^{\{b N_c+\bar n_{\text{max}}-1\}}_{\{\bar n_{\text{max}}-1\}} \right) \\
= & \ \ell\!\left( U[r]\right) + \sum_{\{ \bar n_{\text{max}}\}} k\!\left( \{\bar n_{\text{max}}\} \to \{\bar n_{\text{max}}-1\} \right)  \\
& \quad\quad\quad\quad\quad \times\underbrace{ \left(\sum_{ \{ bN_c+\bar n_{\text{max}} \} } \ell\!\left( r'\right) k\!\left( \{b N_c+\bar n_{\text{max}}\} \to \{b N_c+\bar n_{\text{max}}-1\} \right)\right)}_{=0}  \\
= & \ \ell\!\left( U[r]\right)\ ,
\end{split}
\end{equation}
which verifies Eq.~(\ref{eq:finalstep}).

Third, let us consider PMC$[N_f+1,1]$ relative to irreps $\hat r'_1$ of ${\cal G}[N_f+1,1]$ interpolated by tensors with $bN_c+\bar n_{\text{max}}-1-i^\prime$ upper indices and $\bar n_{\text{max}}-1$ lower indices, whose $U(1)_{H_1}$ charge is $H_1=i^\prime>0$. These irreps appear in the decomposition of tensors with the second highest rank $T^{\{b N_c+\bar n_{\text{max}}-1\}}_{\{\bar n_{\text{max}}-1\}}$ and the highest rank $T^{\{b N_c+\bar n_{\text{max}}\}}_{\{\bar n_{\text{max}}\}}$.  For each such irrep $\hat r'_1$ one has a PMC 
\begin{equation}
\begin{split}
&0=\ell\!\left(\hat r'_1 \right)\\[0.1cm]
& \phantom{0} =\sum_{\{b N_c+\bar n_{\text{max}}-1\}} \ell\!\left(r'\right) k\!\left(\{b N_c+\bar n_{\text{max}}-1\} \to \{b N_c+\bar n_{\text{max}}-1-i^\prime\} \right)\\
&\phantom{0=} +\sum_{\{ \bar n_{\text{max}} \}} k\!\left(\{ \bar n_{\text{max}} \} \to \{\bar n_{\text{max}}-1\}\right) \\
& \phantom{0=+\sum_{\{b N_c+\bar n_{\text{max}}\}}} \times \underbrace{\sum_{\{b N_c+\bar n_{\text{max}}\}} \ell\!\left(r'\right) k\!\left(\{b N_c+\bar n_{\text{max}}\} \to \{b N_c+\bar n_{\text{max}}-1-i^\prime\} \right)}_{=0} \, .
\end{split}
\end{equation}
The last line vanishes because of Eq.~(\ref{eq:typeA_app}); hence we obtain
\begin{equation}
\label{eq:typeA_app_2}
\begin{split}
0 & =  \ell\!\left( \hat r'_1 \right) \\
& = \sum_{\{ bN_c+\bar n_{\text{max}}-1 \} } \ell\!\left( r'\right) k\!\left( \{bN_c+\bar n_{\text{max}}-1\} \to \{bN_c+\bar n_{\text{max}}-1-i^\prime\} \right)\, ,
\end{split}
\end{equation}
which has the same form as Eq.~(\ref{eq:typeA_app}), but the rank of the corresponding interpolating tensors is reduced by one.
We thus compute the sum in Eq.~(\ref{eq:finalstep}) for representations $r$ (with $H_1=0$) interpolated by tensors $T^{\{b N_c+\bar n_{\text{max}}-2\}}_{\{\bar n_{\text{max}}-2\}}$. By making use of Eq.~(\ref{eq:typeA_app}) with $i=2$ and Eq.~(\ref{eq:typeA_app_2}) with $i^\prime=1$ we find
\begin{equation}
\label{eq:typeB_3_app}
\begin{split}
\sum_{r^\prime} \, & \ell(r^\prime ) k\!\left( r^\prime \to r\right) \\
& = \ell\!\left(U[r]\right) +\sum_{\{ \bar n_{\text{max}}\}} k\left(\{\bar n_{\text{max}} \} \to \{\bar n_{\text{max}} -2\} \right)\\
&\phantom{=\ell\!\left(U[r]\right) + \sum_{\{ \bar n_{\text{max}}\}}}
  \times \underbrace{\sum_{\{ b N_c+ \bar n_{\text{max}} \}} \ell\!\left( r'\right) k\left(\{ b N_c+ \bar n_{\text{max}} \} \to \{ b N_c+ \bar n_{\text{max}} -2\} \right)}_{=0}\\
&\phantom{=} +\sum_{\{ \bar n_{\text{max}} -1\}} k\left(\{ \bar n_{\text{max}} -1\} \to \{ \bar n_{\text{max}} -2\} \right)\\
&\phantom{0 = +\sum_{\{n_{\bar q}^{max}\}}} \times \underbrace{\sum_{\{ b N_c+ \bar n_{\text{max}}-1 \}} \ell\!\left( r'\right) k\left(\{ b N_c+ \bar n_{\text{max}}-1\} \to \{b N_c+ \bar n_{\text{max}}-2\} \right)}_{=0} \\
&=\ell\!\left(U[r]\right)\ ,
\end{split}
\end{equation}
which verifies Eq.~(\ref{eq:finalstep}).

We have thus proven Eq.~(\ref{eq:finalstep}) for irreps $r$ interpolated by tensors with the highest rank $T^{\{b N_c+\bar n_{\text{max}}\}}_{\{\bar n_{\text{max}}\}}$, the second highest rank $T^{\{b N_c+\bar n_{\text{max}}-1\}}_{\{\bar n_{\text{max}}-1\}}$, and the third highest rank $T^{\{b N_c+\bar n_{\text{max}}-2\}}_{\{\bar n_{\text{max}}-2\}}$.  By repeating the same steps, one can prove Eq.~(\ref{eq:finalstep}) for irreps interpolated by the tensors of any lower rank.

\bibliography{ChSB_refs}

\end{document}